# ASYMPTOTIC BOUNDS ON THE GLOBALLY OPTIMAL POSITIONS OF ORTHOGONAL STIFFENERS FOR RECTANGULAR PLATES IN ELASTOSTATIC BENDING

Nathan Perchikov*

School of Mechanical Engineering, Faculty of Engineering, Tel Aviv University, Ramat Aviv, Tel Aviv 69978, Israel

**Abstract** The present paper treats the problem of finding the asymptotic bounds for the globally optimal locations of orthogonal stiffeners minimizing the compliance of a rectangular plate in elastostatic bending. The essence of the paper is the utilization of a method of analysis of orthogonally stiffened rectangular plates first presented by Mazurkiewicz in 1962, and obtained herein in a closed form for several special cases in the approximation of stiffeners having zero torsional rigidity. Asymptotic expansions of the expressions for the deflection field of a stiffened plate are used to derive limit-case globally optimal stiffening layouts for highly flexible and highly rigid stiffeners. A central result obtained in this work is an analytical proof of the fact that an array of flexible enough orthogonal stiffeners of any number, stiffening a simply-supported rectangular plate subjected to any lateral loading, is best to be put in the form of exactly two orthogonal stiffeners, one in each direction.

## 1   INTRODUCTION

This paper addresses the well-known problem of optimal positioning of straight, uniform, orthogonal beam-stiffeners of a given number or, integrally attached to a thin, rectangular, isotropic and homogeneous plate subjected to lateral, elastic, quazistatic loading, in a layout minimizing the total work done by the loads, in a formulation resembling an earlier version by Dems et al. (1989). Herein, as in a majority of previous works, it is assumed that the neutral lines of the stiffeners coincide with the neutral surface of the plate.

Both the analysis and the optimal design problems have gained substantial attention in the literature. The early papers of Nowacki (1954a,1954b) and Mazurkiewicz (1962a,1962b) solved the analysis problem by assuming the stiffeners positioned symmetrically about the neutral surface of the plate and enforcing geometric compatibility between the plate and the stiffeners as well as the static equilibrium of both the stiffeners (the Euler-Bernoulli equation) and the plate (the Love-Kirchhoff equation). In the work of Mazurkiewicz (1962), the case of a simply-supported plate was treated and the influence of the stiffeners was viewed as effective distributed line loads expressed as double sine series. The procedure employed by Mazurkiewicz for bidirectional stiffening with beams having non-negligible torsional rigidity produced infinite sets of linear algebraic equations that had to be cut off and solved simultaneously. A proof of the convergence of such algebraic systems is given by Fletcher and Thorne (1955), within the context of the analysis of unstiffened plates having arbitrary boundary conditions. In the work of Dems et al. (1989), the equilibrium equations for the plate and the stiffeners are replaced by the use of the Green functions of the bending of simply-supported plate and beams, and the compatibility equation is formulated as Fredholm's integral equation of the 1$^{st}$ kind, which is solved numerically by assuming that the stiffeners are attached to the plate in a finite number of points and the effective point loads corresponding to geometric compatibility at those points are computed, again by the simultaneous solution of cut infinite linear algebraic equations. Various aspects of the analysis of quazistatic bending, free vibrations and buckling of orthogonally stiffened rectangular simply-supported plates using double sine series representations were examined (Konchkovskii 1984; Savin et al. 1964; Grigolyuk et al. 1980; Andrianov et al. 1985; Kalamkarov 1992).

The (sometimes called inverse) optimization problem, of optimal positioning of stiffeners has gained not less an extensive coverage. In the work of Dems et al. (1989) the locations of the stiffeners are varied to decrease the work done by a given quazistatic lateral loading, where the optimization procedure utilizes an optimality criterion derived variationally. In their work, Grayhack and Mahar (1990) consider highly flexible stiffeners positioned in a layout increasing the minimum buckling load and obtain an analytic form of the equations determining the optimal locations.

The objective of the present paper is the derivation of a closed-form expression for the elastic strain energy in a simply supported thin rectangular plate stiffened with a single set of parallel stiffeners and subjected to a harmonic loading, an asymptotic expansion of the elastic strain energy with respect to a non-dimensional parameter representing the ratio of the bending rigidity of the stiffeners and the plate, and analytical globally optimal minimization of the compliance with respect to the locations of the stiffeners. In addition, a closed form, globally optimal bound for the optimal layout of asymptotically flexible stiffeners is derived for the case of bidirectional stiffening. The essence of the results presented herein is the explicitness, simplicity and global optimality of the expressions for the optimal layouts of highly flexible and highly rigid stiffeners. In the present work, in contrast to the procedure followed by Dems et al. (1989), the geometric compatibility equation (between the plate and the stiffeners) is formulated as Fredholm's $2^{nd}$ kind integral equation, which, unlike the $1^{st}$ kind equation can be solved analytically for the case of a double-sine expressible Green's kernel function. This method is equivalent to the method used by Nowacki (1954) and Mazurkiewicz (1962) with the exception that here a closed form expression is obtained, mainly (but not solely) due to the restriction to unidirectional stiffening. A thorough review of solutions of contact problems in plate and shell bending by reformulation of the governing equations into $2^{nd}$ kind Fredholm's equations can be found in the work of Grigolyuk et al. (1980).

---

* E-mail: perchico@gmail.com

A different approach is described in the books of Andrianov et al. (1980) and Kalamkarov (1992), where the method of homogenization is applied. Finding an anisotropic plate structurally equivalent to a rib-stiffened isotropic plate is widely used in development of finite elements, topological optimization and analysis of uniformly stiffened panels. Although obviously advantageous in the aforementioned cases, the method of homogenization is not necessarily more suitable for the purpose of finding the globally optimal locations of asymptotically rigid stiffeners. A discussion on certain mathematical aspects of the method is given by Andrianov et al. (1980). A review of applications of the method is carried out by Kalamkarov (1992). The outline of the present paper is as follows. In section 2, the deflection field of a stiffened plate is derived self-consistently, assuming continuous interfaces between the plate and the stiffeners, resulting in a closed form expression for a unidirectionally stiffened simply supported plate subjected to bisinusoidal loading, and in double-sine series for general loading cases. In section 3, the solution is generalized to account for bidirectional stiffening. In section 4, torsional rigidity is taken into account for the simpler case of unidirectional stiffening. In section 5, on the basis of the results of sections 2, 3 and 4, asymptotic, explicit and globally optimal bounds for the optimal locations of highly flexible and highly rigid uni- or bidirectional stiffeners with or without torsional rigidity are obtained analytically.

## 2 ANALYSIS OF UNIDIRECTIONALLY STIFFENED PLATES

This section outlines the procedure of derivation of the linear elastic deflection field of a thin, rectangular, homogenous, isotropic, simply supported plate of a uniform thickness, subjected to a prescribed lateral loading and stiffened with straight uniform parallel open-section beams, all perpendicular to one of the sides of the plate, and having locations specified parametrically.

Graphical representation of the problem follows in Fig. 1, where $\xi$ and $\eta$ denote the planar Cartesian coordinates, $a$, $b$ and $t$ the dimensions of the plate, $p(\xi,\eta)$ the distribution of the lateral loading and $x_1$, $x_2$ the locations of the stiffeners. The bending rigidities of the plate and the stiffeners are denoted by $D$, $E_1I_1$ and $E_2I_2$, respectively.

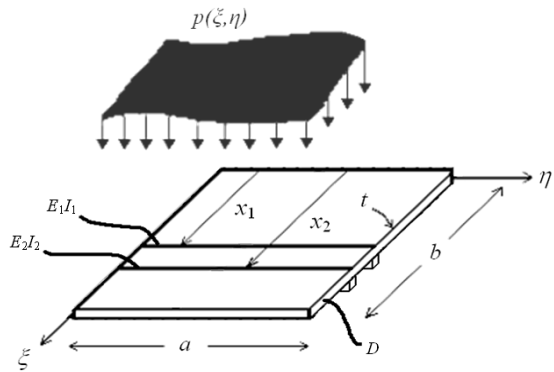

**Fig. 1** Representation of the problem (the lower half of a plate stiffened symmetrically with respect to its neutral surface)

The Love-Kirchhoff model is used to describe the bending of the plate. The stiffeners are assumed to behave as Euler-Bernoulli beams, positioned symmetrically with respect to the neutral surface of the plate and having the same deflection as the neutral surface, along their longitudinal axes. The influence of the $i^{th}$ stiffener on the plate is represented by a distributed force per unit length, $q_i(x_i,\eta)$, defined as positive when acting *upwards*. The stiffeners and the plate are assumed to have the same boundary conditions. In the case considered in the present section the beams are all simply supported. Each beam is assumed to be subjected to a distributed force per unit length $q_i(x_i,\eta)$, defined as positive acting when *downwards*, as shown in Fig. 2.

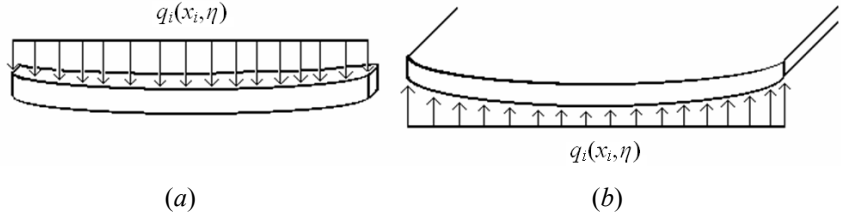

(a)             (b)

**Fig. 2** (*a*) The distributed force that the plate applies on the $i^{th}$ stiffener and (*b*) The distributed force that the $i^{th}$ stiffener applies on the plate

The solution procedure presented here is based on the infinitesimal deformation assumption, according to which the deflection of the plate is small relatively to its thickness, and the thickness, in turn, is small relatively to the (say, geometric) average of the lengths of the edges of the plate. According to this assumption, the bending of the plate is uncoupled with the in-plane stretch and can therefore be described by a single force balance equation for the static equilibrium deflection of the neutral surface (usually the middle surface) of the plate, and the appropriate boundary conditions. This paper treats the problem of small-strain elastic bending of a thin, simply supported plate. This means that the plate is treated as if it had a line of hinges on each of two of its adjacent edges and a line of moving supports, say, moving bearings, on each of the other two edges. Accordingly, the deflection field of the plate would follow from the Love-Kirchhoff assumptions. The derivation of the elastic infinitesimal deflection field will follow from linear superposition:

$$w(\xi,\eta,\mathbf{x}) = w_p(\xi,\eta) + \sum_{i=1}^{N} w_{q_i}(\xi,\eta,\mathbf{x}) = \\ = \int_0^b \int_0^a p(\alpha,\beta) w_1(\xi,\eta,\alpha,\beta) d\beta d\alpha + \sum_{i=1}^{N} \int_0^a q_i(x_i,\beta,\mathbf{x}) w_1(\xi,\eta,x_i,\beta) d\beta \quad (2.1)$$

where $w(\xi,\eta,\mathbf{x})$ is the total deflection of the stiffened plate at a point which coordinates are denoted by $\xi$ and $\eta$, namely, at $(\xi,\eta)$, $w_p(\xi,\eta)$ is the deflection of the plate alone due to the external load alone at $(\xi,\eta)$, $w_1(\xi,\eta,\alpha,\beta)$ is the deflection of the unstiffened plate at $(\xi,\eta)$ due to a unit lateral force acting at $(\alpha,\beta)$, $p(\xi,\eta)$ is the function describing the distribution of the given external lateral load, $\mathbf{x}$ is the vector of the $\xi$ coordinates of the stiffening beams and $\mathbf{q}(\xi,\eta,\mathbf{x})$ is the vector of the distributed forces per unit length representing the forces acting between the plate and the stiffeners (due to the external load) when the system is in equilibrium, each of which components corresponds to each of the $N$ stiffeners.

Equation (2.1) assumes that $p(\xi,\eta)$ is everywhere small enough for the deflection of the plate to be small relatively to its thickness. According to Newton's third law,

$$q_i(x_i,\eta,\mathbf{x}) = -q_i^B(x_i,\eta,\mathbf{x}) \quad (2.2)$$

where the superscript *B* represents the fact that the distributed force is applied at the beams representing the stiffeners. It is now due to *geometric compatibility*, representing the fact that the stiffeners are considered integrally attached to the plate, that the deflection fields of the stiffeners are equal to the deflection field of the stiffened plate, along their interfaces, namely:

$$w(x_i,\eta,\mathbf{x}) = w_i^B(x_i,\eta,\mathbf{x}) \quad (2.3)$$

Next, assuming that the stiffeners can be treated as Euler-Bernoulli beams, one can write:

$$q_i^B(x_i,\eta,\mathbf{x}) = E_i I_i \frac{\partial^4 w_i^B(x_i,\eta,\mathbf{x})}{\partial \eta^4} \quad (2.4)$$

Substituting equations (2.2) and (2.3) into the above relation, results in:

$$q_i(x_i, \eta, \mathbf{x}) = -E_i I_i \frac{\partial^4 w(x_i, \eta, \mathbf{x})}{\partial \eta^4} \equiv -E_i I_i w_\eta^{(IV)}(x_i, \eta, \mathbf{x}) \tag{2.5}$$

Next, substituting equation (2.5) into equation (2.1) produces the integro-differential equation for the deflection field of the stiffened plate (which can be turned into a standard Fredholm integral equation of the 2$^{nd}$ kind – and solved using the standard procedure – either by differentiating or by performing integration by parts four times consecutively):

$$w(\xi, \eta, \mathbf{x}) = \int_0^b \int_0^a p(\alpha, \beta) w_1(\xi, \eta, \alpha, \beta) d\beta d\alpha - \sum_{i=1}^N E_i I_i \int_0^a w_\beta^{(IV)}(x_i, \beta, \mathbf{x}) w_1(\xi, \eta, x_i, \beta) d\beta \tag{2.6}$$

Here, use is made of Green's function for the bending of a Love-Kirchhoff simply-supported plate as obtained by Navier (Szilard (1974)):

$$w_1(\xi, \eta, \alpha, \beta) = \frac{4}{\pi^4 abD} \sum_{n,s=1}^\infty \frac{\sin(n\pi\alpha/b)\sin(s\pi\beta/a)\sin(n\pi\xi/b)\sin(s\pi\eta/a)}{(n^2/b^2 + s^2/a^2)^2} \tag{2.7}$$

and uniform beams are assumed:

$$E_j I_j(\eta) = E_j I_j \quad \forall \ 1 \le j \le N \tag{2.8}$$

Next, taking a double-sine transform of the loading,

$$P_{nr} = \frac{4}{ab} \int_0^b \int_0^a p(\alpha, \beta) \sin(n\pi\alpha/b) \sin(r\pi\beta/a) d\beta d\alpha \tag{2.9}$$

and introducing the auxiliary function:

$$\lambda_{jn} \equiv E_j I_j \sin(n\pi x_j / b) \tag{2.10}$$

the deflection field of a unidirectionally stiffened simply supported plate for a general loading becomes:

$$w(\xi, \eta, \mathbf{x}) = \frac{1}{\pi^4 D} \sum_{m,r=1}^\infty \frac{\sin(m\pi\xi/b)\sin(r\pi\eta/a)}{(m^2/b^2 + r^2/a^2)^2} \left( P_{mr} + \frac{2r^4}{a^4 bD} \left\{ \sum_{n=1}^\infty \frac{\boldsymbol{\lambda}_m^T \sin(n\pi\mathbf{x}/b)}{(n^2/b^2 + r^2/a^2)^2} \left[ \frac{2r^4}{a^4 bD} \times \right.\right.\right.$$
$$\left.\left.\left. \sin(n\pi\mathbf{x}^T/b) \left( 1 + \frac{2r^4}{a^4 bD} \sum_{\mu=1}^\infty \frac{\boldsymbol{\lambda}_\mu \sin(\mu\pi\mathbf{x}^T/b)}{(\mu^2/b^2 + r^2/a^2)^2} \right)^{-1} \sum_{\nu=1}^\infty \frac{\boldsymbol{\lambda}_\nu P_{\nu r}}{(\nu^2/b^2 + r^2/a^2)^2} - P_{nr} \right] \right\} \right) \tag{2.11}$$

The elastic strain energy stored in the deformed structure, consistent of the energy stored in the stiffened plate and the energy stored in the stiffeners, is equal to the work done by the external loads, and can be obtained accordingly as:

$$U = \frac{1}{2} \int_0^a \int_0^b p(\xi, \eta) w(\xi, \eta, \mathbf{x}) d\xi d\eta \tag{2.12}$$

Next, if one assumes a sinusoidal loading distribution,

$$p(\xi, \eta) = P_{gh} \sin(g\pi\xi/b) \sin(h\pi\eta/a) \Rightarrow P_{kr} = P_{gh} \delta_{gk} \delta_{hr} \tag{2.13}$$

The deflection field of the stiffened plate becomes:

$$w(\xi, \eta, \mathbf{x}) = \frac{P_{gh}}{\pi^4 D} \frac{\sin(g\pi\xi/b)\sin(h\pi\eta/a)}{(g^2/b^2 + h^2/a^2)^2} +$$
$$+ \frac{4h^8 P_{gh}}{\pi^4 a^8 b^2 D^3} \sum_{m=1}^\infty \frac{\boldsymbol{\lambda}_m^T \sin(m\pi\xi/b)\sin(h\pi\eta/a)}{(m^2/b^2 + h^2/a^2)^2} \sum_{n=1}^\infty \frac{\sin(n\pi\mathbf{x}/b)}{(n^2/b^2 + h^2/a^2)^2} \times$$
$$\times \sin(n\pi\mathbf{x}^T/b) \left( 1 + \frac{2h^4}{a^4 bD} \sum_{\mu=1}^\infty \frac{\boldsymbol{\lambda}_\mu \sin(\mu\pi\mathbf{x}^T/b)}{(\mu^2/b^2 + h^2/a^2)^2} \right)^{-1} \frac{\boldsymbol{\lambda}_g}{(g^2/b^2 + h^2/a^2)^2}$$
$$- \frac{2h^4 P_{gh}}{\pi^4 a^4 bD^2} \sum_{m=1}^\infty \frac{\boldsymbol{\lambda}_m^T \sin(m\pi\xi/b)\sin(h\pi\eta/a)}{(m^2/b^2 + h^2/a^2)^2} \frac{\sin(g\pi\mathbf{x}/b)}{(g^2/b^2 + h^2/a^2)^2} \tag{2.14}$$

Hence, more compact a form of the deflection field of a stiffened plate can be obtained. First, using a well-known trigonometric identity:

$$\sum_{n=1}^{\infty}\frac{2\sin(n\pi x_j/b)\sin(n\pi x_i/b)}{(n^2/b^2+r^2/a^2)^2}=\sum_{n=1}^{\infty}\frac{\cos[n\pi(x_j-x_i)/b]}{(n^2/b^2+r^2/a^2)^2}-\sum_{n=1}^{\infty}\frac{\cos[n\pi(x_j+x_i)/b]}{(n^2/b^2+r^2/a^2)^2} \quad (2.15)$$

the double sine sums in (2.14) can be written as a difference between two single cosine sums. Then, summing the cosine series analytically by relating them to a function that produces a known sine series, the following result is obtained (this matter is further addressed in section 5.2):

$$\frac{d}{dx}\sum_{n=1}^{\infty}\frac{\cos(n\pi x/b)}{(n^2/b^2+r^2/a^2)^2}=-\sum_{n=1}^{\infty}\frac{(n\pi/b)\sin(n\pi x/b)}{(n^2/b^2+r^2/a^2)^2}=$$
$$=\frac{ab^2\sinh(\pi r x/a)-abx\sinh(\pi r b/a)\cosh(\pi r[b-|x|]/a)}{4\pi^{-3}r\sinh^2(\pi r b/a)} \quad (2.16)$$

By integrating (2.16) and substituting the result into (2.14), an expression with no infinite sums is obtained. Defining, component-wise, the following quantities (*i* and *j* being the indices related to the numbers of the stiffeners):

$$B_{jir}=\frac{r^2 E_j I_j}{2\pi^3 a^2 D}\left\{b\frac{\cosh(\pi r[x_j-x_i]/a)-\cosh(\pi r[x_j+x_i]/a)}{\sinh^2(\pi r b/a)}+a\frac{\cosh(\pi r[b-|x_j-x_i|]/a)}{\pi r\sinh(\pi r b/a)}\right.$$
$$\left.-a\frac{\cosh(\pi r[b-x_i-x_j]/a)}{\pi r\sinh(\pi r b/a)}+\frac{|x_j-x_i|\sinh(\pi r[b-|x_j-x_i|]/a)-(x_j+x_i)\sinh(\pi r[b-x_i-x_j]/a)}{\sinh(\pi r b/a)}\right\} \quad (2.17)$$

$$s_{jh}=\frac{P_{gh}h^2\sin(h\pi\eta/a)E_j I_j}{2\pi^3 a^2 D}\left\{b\frac{\cosh(\pi h[x_j-\xi]/a)-\cosh(\pi h[x_j+\xi]/a)}{\sinh^2(\pi h b/a)}+a\frac{\cosh(\pi h[b-|x_j-\xi|]/a)}{\pi h\sinh(\pi h b/a)}\right.$$
$$\left.-a\frac{\cosh(\pi h[b-\xi-x_j]/a)}{\pi h\sinh(\pi h b/a)}+\frac{|x_j-\xi|\sinh(\pi h[b-|x_j-\xi|]/a)-(x_j+\xi_0)\sinh(\pi h[b-\xi-x_j]/a)}{\sinh(\pi h b/a)}\right\} \quad (2.18)$$

$$S_{jih}=\frac{h^2}{2\pi^3 a^2 D}\left\{b\frac{\cosh(\pi h[x_j-x_i]/a)-\cosh(\pi h[x_j+x_i]/a)}{\sinh^2(\pi h b/a)}+a\frac{\cosh(\pi h[b-|x_j-x_i|]/a)}{\pi h\sinh(\pi r b/a)}\right.$$
$$\left.-a\frac{\cosh(\pi h[b-x_i-x_j]/a)}{\pi h\sinh(\pi r b/a)}+\frac{|x_j-x_i|\sinh(\pi h[b-|x_j-x_i|]/a)-(x_j+x_i)\sinh(\pi h[b-x_i-x_j]/a)}{\sinh(\pi h b/a)}\right\} \quad (2.19)$$

the deflection field for a bisinusoidal loading in a closed form – although not totally explicitly, as a solution of a finite linear system of equations is required – becomes:

$$w_{gh}(\xi,\eta,\mathbf{x})=\frac{P_{gh}\sin(g\pi\xi/b)\sin(h\pi\eta/a)+\mathbf{s}_h^T[\mathbf{S}_h(\mathbf{1}+\mathbf{B}_h)^{-1}\boldsymbol{\lambda}_g-\sin(g\pi\mathbf{x}/b)]}{\pi^4 D(g^2/b^2+h^2/a^2)^2} \quad (2.20)$$

Having the deflection field corresponding to a bisinusoidal loading, the deflection field for a general loading (expressible as a double sine series) can be derived by superposition:

$$w(\xi,\eta,\mathbf{x})=\sum_{g,h=1}^{\infty}\frac{P_{gh}\sin(g\pi\xi/b)\sin(h\pi\eta/a)+\mathbf{s}_h^T[\mathbf{S}_h(\mathbf{1}+\mathbf{B}_h)^{-1}\boldsymbol{\lambda}_g-\sin(g\pi\mathbf{x}/b)]}{\pi^4 D(g^2/b^2+h^2/a^2)^2} \quad (2.21)$$

Numerical verification of the presented method is omitted here, as an equivalent method has been presented before by Nowacki (1954) and Mazurkiewicz (1962). The analytical summations performed here can only improve convergence and help avoid the Gibbs phenomena.

However, in the present paper, the above operations are needed in order to obtain closed-form asymptotic expansions rather than in order to produce a better numerical analysis module. Therefore, no comparison of deflections and stresses of stiffened plates for various loading cases (calculated once with the employed method and once with commercial software tools), is given here.

# 3 GENERALIZATION OF THE ANALYSIS FOR BIDIRECTIONAL STIFFENING

The treatment of bidirectional stiffening is not much different from that of the unidirectional one. The deflection field can be obtained using linear superposition:

$$w(\xi,\eta,\mathbf{x}) = w_p(\xi,\eta) + \sum_{i=1}^{N} w_{q_i}(\xi,\eta,\mathbf{x}) = \int_0^b \int_0^a p(\alpha,\beta) w_1(\xi,\eta,\alpha,\beta) d\beta d\alpha +$$
$$+ \sum_{i=1}^{N_\eta} \int_0^a q_i(x_i,\beta,\mathbf{x}) w_1(\xi,\eta,x_i,\beta) d\beta + \sum_{i=N_\eta+1}^{N_\eta+N_\xi} \int_0^b q_i(\alpha,x_i,\mathbf{x}) w_1(\xi,\eta,\alpha,x_i) d\alpha \quad (3.1)$$

where $q_i$ are the distributed forces per unit length exerted on the plate by the stiffeners, $N$ is the total number of the stiffeners, $N_\eta$ is the number of $\eta$-axis aligned stiffeners, and $N_\xi$ is the number of $\xi$-axis aligned stiffeners. Denoting the interaction forces applied on the $\xi$-axis aligned stiffeners by the $\eta$-axis aligned stiffeners by $f_{ij}$, one can write the following relations:

$$E_i I_i \frac{\partial^4 w(\xi,x_i)}{\partial \xi^4} = E_i I_i \frac{\partial^4 w_{i's\,Beam}(\xi)}{\partial \xi^4} = q_{on\,i's\,Beam}(\xi) = -q_i(\xi,x_i) + \sum_{j=1}^{N_\eta} f_{ij} \delta(\xi - x_j) \quad (3.2)$$

where the first equality sign is due to the assumption that the neutral axes of the stiffeners coincide with the neutral surface of the plate, the second is due to the Euler-Bernoulli beam theory for uniform beams and the third is obtained by summing up all the forces exerted on the $i$'s, $\xi$-axis aligned stiffener. The first term after the third equality sign is the distributed force per unit length exerted by the plate on the $i$'s, $\xi$-axis aligned stiffener, as arises from Newton's third law. The function $\delta(x)$ denotes Dirac's Delta function. For the $\eta$-axis aligned stiffeners one can write:

$$E_i I_i \frac{\partial^4 w(x_i,\eta)}{\partial \eta^4} = -q_i(x_i,\eta) + \sum_{j=N_\eta+1}^{N_\eta+N_\xi} (-f_{ij}) \delta(\eta - x_j) \quad (3.3)$$

using the same arguments as before and having $(-f_{ij})$ instead of $f_{ij}$ this time, due to Newton's third law.

Next, deriving $q_i$ from (3.2) and (3.3) and substituting them into (3.1), one has:

$$w(\xi,\eta,\mathbf{x}) = \int_0^b \int_0^a p(\alpha,\beta) w_1(\xi,\eta,\alpha,\beta) d\beta d\alpha - \sum_{i=1}^{N_\xi} \int_0^a E_i I_i w_\beta^{(IV)}(x_i,\beta,\mathbf{x}) w_1(\xi,\eta,x_i,\beta) d\beta$$
$$- \sum_{i=1}^{N_\eta} \int_0^a \sum_{i=N_\eta+1}^{N_\eta+N_\xi} f_{ij} \delta(\beta - x_j) w_1(\xi,\eta,x_i,\beta) d\beta - \sum_{i=N_\eta+1}^{N_\eta+N_\xi} \int_0^b E_i I_i w_\alpha^{(IV)}(\alpha,x_i,\mathbf{x}) w_1(\xi,\eta,\alpha,x_i) d\alpha +$$
$$+ \sum_{i=N_\eta+1}^{N_\eta+N_\xi} \int_0^b \sum_{i=1}^{N_\eta} f_{ij} \delta(\alpha - x_j) w_1(\xi,\eta,\alpha,x_i) d\alpha = \int_0^b \int_0^a p(\alpha,\beta) w_1(\xi,\eta,\alpha,\beta) d\beta d\alpha \quad (3.4)$$
$$- \sum_{i=1}^{N_\xi} \int_0^a E_i I_i w_\beta^{(IV)}(x_i,\beta,\mathbf{x}) w_1(\xi,\eta,x_i,\beta) d\beta - \sum_{i=N_\eta+1}^{N_\eta+N_\xi} \int_0^b E_i I_i w_\alpha^{(IV)}(\alpha,x_i,\mathbf{x}) w_1(\xi,\eta,\alpha,x_i) d\alpha$$
$$+ \sum_{i=N_\eta+1}^{N_\eta+N_\xi} \sum_{i=1}^{N_\eta} f_{ij} w_1(\xi,\eta,x_j,x_i) - \sum_{i=1}^{N_\eta} \sum_{i=N_\eta+1}^{N_\eta+N_\xi} f_{ij} w_1(\xi,\eta,x_i,x_j)$$

Owing to the fact that the influence function for the bending of a simply-supported plate, $w_1$, is symmetric (insensitive to an interchange of the arguments), as one can learn from (2.7), the last two terms in (3.4) cancel each other. Therefore, in deriving the compatibility equation for the bending of an orthogonally stiffened simply-supported plate, one can disregard the interaction forces between the stiffeners, as they cancel out in the final governing equation. Consequently, the governing equation becomes:

$$w(\xi,\eta,\mathbf{x}) = \int_0^b \int_0^a p(\alpha,\beta) w_1(\xi,\eta,\alpha,\beta) d\beta d\alpha - \sum_{i=1}^{N_\xi} \int_0^a E_i I_i w_\beta^{(IV)}(x_i,\beta,\mathbf{x}) w_1(\xi,\eta,x_i,\beta) d\beta$$
$$- \sum_{i=N_\eta+1}^{N_\eta+N_\xi} \int_0^b E_i I_i w_\alpha^{(IV)}(\alpha,x_i,\mathbf{x}) w_1(\xi,\eta,\alpha,x_i) d\alpha \quad (3.5)$$

Hence, using expressions (2.7)-(2.10) and the orthogonality of the sine functions, and following the strategy for the solution of integro-differential equations mentioned in section 2, one obtains two sets of compatibility equations for the distributed forces between the stiffeners and the plate, one for each stiffening direction.

Making use of (2.17) and introducing the following quantities:

$$z_{in} \equiv E_i I_i \int_0^b w_\alpha^{(IV)}(\alpha, x_i, \mathbf{x}) \sin(n\pi\alpha/b) d\alpha \ ; \ v_{is} \equiv E_i I_i \int_0^a w_\beta^{(IV)}(x_i, \beta, \mathbf{x}) \sin(s\pi\beta/a) d\beta$$

$$u_{jr} \equiv \frac{r^4}{2a^3 D} \sum_{n=1}^\infty \frac{\lambda_{jn} P_{nr}}{(n^2/b^2 + r^2/a^2)^2} \ ; \ y_{jm} \equiv \frac{bm^4}{2a^4 D} \sum_{s=1}^\infty \frac{\mu_{js} P_{ms}}{(m^2/b^2 + s^2/a^2)^2} \ ; \ \mu_{js} \equiv E_j I_j \sin(s\pi x_j/a) \quad (3.6)$$

$$C_{jirn} \equiv \frac{2r^4 \lambda_{jn} \sin(r\pi x_i/a)}{a^4 b D(n^2/b^2 + r^2/a^2)^2} ; D_{jims} \equiv \frac{2m^4 \mu_{js} \sin(m\pi x_i/b)}{a^5 D(m^2/b^2 + s^2/a^2)^2} ; F_{jim} \equiv \sum_{s=1}^\infty \frac{2m^4 \mu_{js} \sin(s\pi x_i/a)}{a^5 D(m^2/b^2 + s^2/a^2)^2}$$

and (analytically) decoupling the two sets of compatibility equations mentioned above, one gets a single system defined by the quantities:

$$\mathbf{g}_n \equiv (\mathbf{1} + \mathbf{F}_n)^{-1} \left[ \mathbf{y}_n - \sum_{s=1}^\infty \mathbf{D}_{ns} (\mathbf{1} + \mathbf{B}_s)^{-1} \mathbf{u}_s \right]$$

$$\mathbf{M}_{nm} \equiv \sum_{s=1}^\infty (\mathbf{1} + \mathbf{F}_n)^{-1} \mathbf{D}_{ns} (\mathbf{1} + \mathbf{B}_s)^{-1} \mathbf{C}_{sm} \quad (3.7)$$

which are, respectively, the vector and matrix infinite sets representing the final form of the compatibility equation for the bidirectionally stiffened simply supported thin rectangular plate:

$$\mathbf{z}_n = \mathbf{g}_n + \sum_{m=1}^\infty \mathbf{M}_{nm} \mathbf{z}_m \quad (3.8)$$

Equation (3.8) is an infinite inhomogeneous linear system of equations in which every equation is by itself a finite system of equations where the number of equations is equal to the number of the stiffeners in the direction parallel to the $\xi$ axis.

Obviously the relation $\mathbf{M}_{nm} = \mathbf{M}_n \delta_{mn}$ does not hold for any $\mathbf{M}_n$ and therefore the infinite system cannot be solved exactly and one has to truncate it in order to solve it. Thus, the system that would have to be solved would be:

$$\mathbf{z}_n = \mathbf{g}_n + \sum_{m=1}^{m_f} \mathbf{M}_{nm} \mathbf{z}_m \ ; \ 1 \leq n \leq m_f \ll \infty \quad (3.9)$$

In order to solve the system, it is necessary to write the finite systems of equations one beneath the other, resulting in a total number of $m_{Tot} = N_y m_f$ unknowns and equations. That is if one has 17 stiffeners in the direction parallel to the $\xi$ axis and they taking 100 harmonics in the Fourier series representation of the Green influence function of the bending of the plate, then one would write the first 17 equations for $m = 1$, and then the 17 equations for $m = 2$ and so forth up to $m = 100$, so that eventually they would have 1700 equations or rows in the representing matrix. Similarly, in the vectors one would have 17 components for $m = 1$ and then 17 components for $m = 2$ and so forth, with resultant vectors of 1700 components, corresponding to 1700 columns in the representative matrix.

Naturally, one could have written the 100 equations corresponding to the 1st stiffener first and have 17 bunches of 100 and not 100 bunches of 17. However, it is better to keep the equations corresponding to the different stiffeners in bunches so that the truncation error will affect them as similarly as possible, so that no stiffener is represented worse than the other.

Of course, there still remains the order within each bunch, but this is a smaller effect, there has to be some order anyway. Now, if the solution of the system of equations is achieved by Gauss elimination, then the last equations will have the smallest round-off error as the first to be solved, so that it would be beneficiary to put the first, more important harmonics in the end, that is, to reverse the order described above.

Anyhow, defining:

$$\overline{\mathbf{z}} \equiv \begin{pmatrix} \mathbf{z}_{m_f} \\ \mathbf{z}_{m_f-1} \\ \mathbf{z}_{m_f-2} \\ . \\ . \\ . \\ \mathbf{z}_1 \end{pmatrix} \; ; \; \overline{\mathbf{g}} \equiv \begin{pmatrix} \mathbf{g}_{m_f} \\ \mathbf{g}_{m_f-1} \\ \mathbf{g}_{m_f-2} \\ . \\ . \\ . \\ \mathbf{g}_1 \end{pmatrix} \; ; \; \overline{\mathbf{M}} \equiv \begin{bmatrix} \mathbf{M}_{m_f,m_f} & \mathbf{M}_{m_f,m_f-1} & \mathbf{M}_{m_f,m_f-2} & . & . & \mathbf{M}_{m_f,1} \\ \mathbf{M}_{m_f-1,m_f} & \mathbf{M}_{m_f-1,m_f-1} & \mathbf{M}_{m_f-1,m_f-2} & . & . & \mathbf{M}_{m_f-1,1} \\ \mathbf{M}_{m_f-2,m_f} & \mathbf{M}_{m_f-2,m_f-1} & . & . & . & . \\ . & . & . & . & . & . \\ . & . & . & . & . & . \\ \mathbf{M}_{1,m_f} & \mathbf{M}_{1,m_f-1} & . & . & . & \mathbf{M}_{1,1} \end{bmatrix} \quad (3.10)$$

every component in the vectors being a column vector itself and every component in the matrix being a matrix itself, results in the equation:

$$\overline{\mathbf{z}} = \overline{\mathbf{g}} + \overline{\mathbf{M}}\overline{\mathbf{z}} \Rightarrow (\mathbf{1} - \overline{\mathbf{M}})\overline{\mathbf{z}} = \overline{\mathbf{g}} \qquad (3.11)$$

where the unit matrix is $\mathbf{1} \in \mathbb{R}^{m_f \times m_f}$.

The solution of equation (3.8) would produce a series of the approximate values of a finite number of the necessary integrals for the evaluation of the deflection field. If the matrix $(\mathbf{1} - \overline{\mathbf{M}})$ is regular and the system is convergent, then the solution can be written as:

$$\overline{\mathbf{z}} = (\mathbf{1} - \overline{\mathbf{M}})^{-1}\overline{\mathbf{g}} \qquad (3.12)$$

Obviously, this is only formal writing. The solution of a system of equations is not achieved by taking the inverse of a large matrix. The system can be solved with the Matlab solver based on LU factorization and Gauss elimination.

Next, the question of convergence can be addressed in a two-fold way. First, a proof by Fletcher and Thorne [2] guarantees that a system equivalent to (3.8)-(3.11) is convergent and may be solved to any desired degree of accuracy (unlike, say, an asymptotic system, which can only be solved to a certain, limited degree of accuracy).

Second, the system in (3.11) was solved by the author for realistic problem parameters, showing monotonic convergence upon increase in the number of considered harmonics and correlating well with a FEM solution.

Now, whereas the solution of a unidirectionally stiffened plate presented in section 2 can be considered analytical and exact, the solution of a bidirectionally stiffened plate may be considered exact and semi-analytical. 'Exact' due to the convergence of the above series, since exactness of a solution is one's ability to have it up to any desired degree, which is possible in the present case and 'semi-analytic' due to the truncation of the infinite series.

The procedure one has to follow in order to obtain a solution after (3.12) is not discussed thoroughly here. Shortly, one has to solve (3.11)-(3.12) and substitute the solution into (3.5) using definitions (3.10) and (3.6) and (2.7), thus obtaining the third row in (3.5).

For the second row in (3.5) one has to substitute the solution of (3.11) along with the definitions in (3.6) into the following relation, obtained from the analytical decoupling of (3.5):

$$\mathbf{v}_r = (\mathbf{1} + \mathbf{B}_r)^{-1}\left(\mathbf{u}_r - \sum_{n=1}^{m_f} \mathbf{C}_{rn}\mathbf{z}_n\right) \qquad (3.13)$$

## 4   UNIDIRECTIONAL STIFFENERS WITH NON-NEGLIGIBLE TORSIONAL RIGIDITY

One can account for the torsional rigidities of the stiffeners (which are relevant especially for the case of stiffeners having closed cross-sections). The torques that the stiffeners apply on the plate can be related to the twisting angles of the plate along the plate-stiffeners interfaces, as follows: for each stiffener, one can employ the constitutional relation of a linear elastic shaft, denoting the torque applied on the plate by $T$, the shear modulus and the geometric torsion rigidity parameter by $G$ and $C$ respectively, the coordinate changing along the stiffener being $\eta$ and the perpendicular being $\xi$, assuming small twisting angles. The superscript $b$ implies that a quantity is related to the stiffener.

$$T_i(x_i,\eta,\mathbf{x}) \underset{\substack{\text{Newton}\\3^{rd}\,Law}}{=} -T_i^b(x_i,\eta,\mathbf{x}) \underset{\substack{\text{Shaft}\\Law}}{=} -C_iG_i\frac{\partial \theta_i^b}{\partial \eta}\underset{w^b=w}{=} -C_iG_i\frac{\partial \theta_i}{\partial \eta}\underset{\theta_i<<1}{=} -C_iG_iw''_{\xi\eta}(x_i,\eta,\mathbf{x}) \qquad (4.1)$$

Superimposing, the deflection field of the (for simplicity, unidirectionally) stiffened plate becomes:

$$w(\xi,\eta,\mathbf{x}) = w_p(\xi,\eta) + \sum_{i=1}^{N} w_{q_i}(\xi,\eta,\mathbf{x}) + \sum_{i=1}^{N} w_{T_i}(\xi,\eta,\mathbf{x}) = \int_0^b \int_0^a p(\alpha,\beta) w_1(\xi,\eta,\alpha,\beta) d\beta d\alpha +$$
$$+ \sum_{i=1}^{N} \int_0^a q_i(x_i,\beta,\mathbf{x}) w_1(\xi,\eta,x_i,\beta) d\beta + \sum_{i=1}^{N} \int_0^a T_i(x_i,\beta,\mathbf{x}) w_{T=1}(\xi,\eta,x_i,\beta) d\beta \quad (4.2)$$

Substituting (4.1) and (2.5) into (4.2) one gets the following integro-differential equation for the deflection field, where only the deflection due to a unit couple, $w_{T=1}$, is yet to be determined:

$$w(\xi,\eta,\mathbf{x}) = \int_0^b \int_0^a p(\alpha,\beta) w_1(\xi,\eta,\alpha,\beta) d\beta d\alpha - \sum_{i=1}^{N} E_i I_i \int_0^a w_\beta^{(IV)}(x_i,\beta,\mathbf{x}) w_1(\xi,\eta,x_i,\beta) d\beta$$
$$- \sum_{i=1}^{N} G_i C_i \int_0^a w_{\xi\beta}''(x_i,\beta,\mathbf{x}) w_{T=1}(\xi,\eta,x_i,\beta) d\beta \quad (4.3)$$

By Betti's theorem, the deflection of an elastic structure due to a unit couple at a point is equal to the rotation angle (in the direction of the couple – perpendicular to the stiffeners, in the present case) – or the slope, for small rotation angles – due to a unit force at that point, namely:

$$w_{T=1}(\xi,\eta,x_i,\beta) = \left.\frac{\partial w_1(\xi,\eta,\alpha,\beta)}{\partial \alpha}\right|_{\alpha=x_i} \quad (4.4)$$

Substituting (4.4) into (2.7) this yields:

$$w_{T=1}(\xi,\eta,x_i,\beta) = \frac{4}{\pi^3 ab^2 D} \sum_{n,s=1}^{\infty} \frac{n \cos(n\pi x_i/b) \sin(s\pi\beta/a) \sin(n\pi\xi/b) \sin(s\pi\eta/a)}{(n^2/b^2 + s^2/a^2)^2} \quad (4.5)$$

Substituting (2.7) and (4.5) into (4.3) produces a fully defined integro-differential equation for the deflection field. Naturally, similar steps can be taken if a second, perpendicular, stiffening array is present.

Treating the obtained integro-differential equation in a manner similar to the one described in the previous sections, and introducing the following quantities (referring to definition (2.10) again):

$$\tau_{jn} \equiv G_j C_j \cos(n\pi x_j/b) \quad , \quad \mathbf{T}_r \equiv \frac{2\pi r^4}{a^4 b^2 D} \sum_{n=1}^{\infty} \frac{n \boldsymbol{\lambda}_n \cos(n\pi \mathbf{x}^T/b)}{(n^2/b^2 + r^2/a^2)^2} \quad (4.6)$$

one obtains a first set of compatibility equations:

$$\mathbf{v}_r = \mathbf{u}_r - \mathbf{B}_r \mathbf{v}_r - \mathbf{T}_r \mathbf{t}_r \quad (4.7)$$

where use is made of the definitions of $\mathbf{B}_r$ from (2.17) and $\mathbf{u}_r$ and $\mathbf{v}_r$ from (3.6) and:

$$t_{is} \equiv G_i C_i \int_0^a w_{\xi\beta}''(x_i,\beta,\mathbf{x}) \sin(s\pi\beta/a) d\beta \quad (4.8)$$

The equation in (4.7) is obtained by calculating the distributed forces-induced deflections at the plate-stiffeners interfaces from the left-hand side of (4.3). Calculating the distributed torques-induced deflections at the plate-stiffeners interfaces from (4.3) results in a second set of compatibility equations:

$$\mathbf{t}_r = \boldsymbol{\upsilon}_r - \sum_{s=1}^{\infty} \boldsymbol{\Lambda}_{rs} \mathbf{v}_s - \sum_{s=1}^{\infty} \boldsymbol{\Omega}_{rs} \mathbf{t}_s \quad (4.9)$$

where the following definitions have been made:

$$\boldsymbol{\upsilon}_r \equiv \frac{1}{\pi^3 bD} \sum_{s=1}^{\infty} \frac{rs - (-1)^{r+s} s^2}{r^2 - s^2} \sum_{n=1}^{\infty} \frac{n \boldsymbol{\tau}_n P_{ns}}{(n^2/b^2 + s^2/a^2)^2}$$
$$\boldsymbol{\Lambda}_{rs} \equiv \frac{4}{\pi^3 ab^2 D} \frac{rs - (-1)^{r+s} s^2}{r^2 - s^2} \sum_{n=1}^{\infty} \frac{n \boldsymbol{\tau}_n \sin(n\pi \mathbf{x}^T/b)}{(n^2/b^2 + s^2/a^2)^2} \quad (4.10)$$
$$\boldsymbol{\Omega}_{rs} \equiv \frac{4}{\pi^2 ab^3 D} \frac{rs - (-1)^{r+s} s^2}{r^2 - s^2} \sum_{n=1}^{\infty} \frac{n^2 \boldsymbol{\tau}_n \cos(n\pi \mathbf{x}^T/b)}{(n^2/b^2 + s^2/a^2)^2}$$

One can decouple the two systems (4.7) and (4.9) analytically.

Defining the following auxiliary quantities:

$$\Psi_{rs} \equiv \Omega_{rs} - \Lambda_{rs}(1+B_s)^{-1}T_s + \delta_{rs}\mathbf{1} \ , \ \varphi_r \equiv \upsilon_r - \sum_{s=1}^{\infty}\Lambda_{rs}(1+B_s)^{-1}u_s \qquad (4.11)$$

where $\delta_{rs}$ is Kronecker's Delta, the following infinite set of equations is obtained for $\mathbf{t}_r$:

$$\sum_{s=1}^{\infty}\Psi_{rs}\mathbf{t}_s = \varphi_r \qquad (4.12)$$

Now, naturally, in order to be solved, this system has to be truncated. If the system is convergent, and we assume that it is, as we had assumed for the other systems in this paper (relying on a proof of convergence by Fletcher and Thorne (1955)), then an approximate solution with any desired degree of accuracy, can be obtained from:

$$\sum_{s=1}^{s_f}\Psi_{rs}\mathbf{t}_s = \varphi_r \ ; 1 \leq r \leq s_f << \infty \qquad (4.13)$$

Next, applying the argumentation from section 3 and denoting by $\overline{\mathbf{t}}$ the truncated vector $(\mathbf{t}_1^T,\mathbf{t}_2^T,...,\mathbf{t}_{s_f}^T)^T$ and doing the same for the other quantities in (4.13), one can write the last form of the compatibility equation for a rectangular, simply supported plate stiffened with a single set of uniform stiffeners, parallel to one of the sides of the plate, taking the torsional rigidity of the stiffeners into account, as follows:

$$\overline{\Psi}\overline{\mathbf{t}} = \overline{\varphi} \qquad (4.14)$$

This system would have to be solved numerically, after defining the auxiliary quantities introduced above. Hence,

$$\theta_r \equiv (1+B_r)^{-1}u_r \ ; \ \Theta_r \equiv (1+B_r)^{-1}T_r \qquad (4.15)$$

and denoting by a corresponding symbol but with an upper bar the truncated vectors $(\mathbf{v}_1^T,\mathbf{v}_2^T,...,\mathbf{v}_{s_f}^T)^T$ and $(\theta_1^T,\theta_2^T,...,\theta_{s_f}^T)^T$ and constructing the truncated matrix $\overline{\Theta}$ in a manner similar to the one employed in equation (3.10) in section 3, we obtain the second set of unknowns required for the solution:

$$\overline{\mathbf{v}} = \overline{\theta} - \overline{\Theta}\overline{\mathbf{t}} \qquad (4.16)$$

And the solution becomes:

$$w(\xi,\eta,\mathbf{x}) = \frac{4}{\pi^4 abD}\sum_{n,s=1}^{s_f}\frac{\sin(n\pi\xi/b)\sin(s\pi\eta/a)}{(n^2/b^2+s^2/a^2)^2}\left[\frac{ab}{4}P_{ns} - \sin(n\pi\mathbf{x}^T/b)\mathbf{v}_s - \frac{n\pi}{b}\cos(n\pi\mathbf{x}^T/b)\mathbf{t}_s\right] \qquad (4.17)$$

Last, in order to be substituted into (4.17), the solutions of (4.14) and (4.16) should be separated into the appropriate sub-vectors again, in a way similar to the way they were constructed.

## 5 ASYMPTOTIC ANALYSIS: GLOBALLY OPTIMAL LAYOUTS OF HIGHLY FLEXIBLE AND HIGHLY RIGID STIFFENERS

This section is concerned with the determination of the optimal layout of stiffeners in the limit case where the stiffeners are much less or much more rigid than the ground structure.

The examination of this limit case is important for the estimation of the correct asymptotic behavior of the optimal locations of stiffeners with respect to their rigidities and the determination of the bounds for the optimal positions of realistic stiffeners. The derivation assumes a thin, isotropic, homogenous, rectangular plate in elastostatic bending, subjected to a lateral pressure and stiffened with two sets of straight, parallel uniform stiffeners.

## 5.1 Highly flexible bidirectional stiffeners with no torsional rigidity – 1st order asymptotics

The linear elastostatic strain energy for the case can be derived, using the results of sections 2 and 3, as follows. Introducing the small parameter $\varepsilon \equiv EI/(aD)$ and deriving the auxiliary quantities required for the calculation of the strain energy for the asymptotic case, one gets:

$$\mathbf{g}_n \underset{(3.7)}{=} (1 + \underbrace{\mathbf{F}_n}_{O(\varepsilon)})^{-1} \left[ \mathbf{y}_n - \sum_{s=1}^{\infty} \mathbf{D}_{ns} (1 + \underbrace{\mathbf{B}_s}_{O(\varepsilon)})^{-1} \mathbf{u}_s \right] \underset{\varepsilon \ll 1}{\to} \underbrace{\mathbf{y}_n}_{O(\varepsilon)} - \sum_{s=1}^{\infty} \underbrace{\underbrace{\mathbf{D}_{ns}}_{O(\varepsilon)} \underbrace{\mathbf{u}_s}_{O(\varepsilon)}}_{O(\varepsilon^2)} \underset{\varepsilon \ll 1}{\to} \underbrace{\mathbf{y}_n}_{O(\varepsilon)} \quad (5.1.1)$$

$$\mathbf{M}_{nm} \underset{(3.7)}{=} \sum_{s=1}^{\infty} (1 + \underbrace{\mathbf{F}_n}_{O(\varepsilon)})^{-1} \mathbf{D}_{ns} (1 + \underbrace{\mathbf{B}_s}_{O(\varepsilon)})^{-1} \mathbf{C}_{sm} \underset{\varepsilon \ll 1}{\to} \sum_{s=1}^{\infty} \underbrace{\mathbf{D}_{ns}}_{O(\varepsilon)} \underbrace{\mathbf{C}_{sm}}_{O(\varepsilon)} = O(\varepsilon^2)$$

$$\Rightarrow \overline{\mathbf{z}} \underset{(3.9)}{=} (1 - \underbrace{\overline{\mathbf{M}}}_{O(\varepsilon^2)})^{-1} \overline{\mathbf{g}} \underset{\varepsilon \ll 1}{\to} \overline{\mathbf{g}} \underset{\varepsilon \ll 1}{\to} \overline{\mathbf{y}} \Rightarrow \underline{\mathbf{z}_n \underset{\varepsilon \ll 1}{\to} \mathbf{y}_n} \quad (5.1.2)$$

$$\mathbf{v}_r \underset{(3.13)}{=} (1 + \underbrace{\mathbf{B}_r}_{O(\varepsilon)})^{-1} (\mathbf{u}_r - \sum_{n=1}^{m_f} \mathbf{C}_{rn} \mathbf{z}_n) \underset{\varepsilon \ll 1}{\to} \mathbf{u}_r - \sum_{n=1}^{m_f} \underbrace{\mathbf{C}_{rn} \mathbf{z}_n}_{\varepsilon \ll 1} \underset{O(\varepsilon)}{\to} \mathbf{u}_r - \sum_{n=1}^{m_f} \underbrace{\underbrace{\mathbf{C}_{rn}}_{O(\varepsilon)} \underbrace{\mathbf{y}_n}_{O(\varepsilon)}}_{\varepsilon \ll 1} \to \mathbf{u}_r \Rightarrow \mathbf{v}_r \underset{\varepsilon \ll 1}{\to} \mathbf{u}_r \quad (5.1.3)$$

Recalling the expression for the deflection (which is not given explicitly in section 3 but can easily be derived):

$$w(\xi, \eta, \mathbf{x}) = \frac{4}{\pi^4 abD} \sum_{n,s=1}^{m_f} \frac{\sin(n\pi\xi/b) in(s\pi\eta/a)}{(n^2/b^2 + s^2/a^2)^2} \left[ \frac{abP_{ns}}{4} - \sum_{i=1}^{N_\eta} \sin(n\pi x_i/b) v_{is} - \sum_{i=N_\eta+1}^{N_\eta+N_\xi} \sin(s\pi x_i/a) z_{in} \right] \quad (5.1.4)$$

integrating it multiplied by half the loading, thus obtaining an expression for the work or the strain energy,

$$U(\mathbf{x}, N_\xi, N_\eta) = \frac{1}{2\pi^4 D} \sum_{n,s=1}^{m_f} \frac{P_{ns}}{(n^2/b^2 + s^2/a^2)^2} \left[ \frac{abP_{ns}}{4} - \sum_{i=1}^{N_\eta} \sin(n\pi x_i/b) v_{is} - \sum_{i=N_\eta+1}^{N_\eta+N_\xi} \sin(s\pi x_i/a) z_{in} \right] \quad (5.1.5)$$

and substituting (5.1.3) into (5.1.5), one gets an explicit expression for the strain energy of the stiffened plate:

$$U(\mathbf{x}, N_\xi, N_\eta) \underset{\min_j \{E_j I_j\} \ll aD}{\to} \frac{1}{2\pi^4 D} \sum_{n,s=1}^{m_f} \frac{P_{ns}}{(n^2/b^2 + s^2/a^2)^2} \left[ \frac{abP_{ns}}{4} - \sum_{i=1}^{N_\eta} \sin(n\pi x_i/b) u_{is} - \sum_{i=N_\eta+1}^{N_\eta+N_\xi} \sin(s\pi x_i/a) y_{in} \right]$$

$$= \frac{1}{2\pi^4 D} \sum_{n,s=1}^{m_f} \frac{abP_{ns}^2}{4(n^2/b^2 + s^2/a^2)^2} - \frac{1}{2\pi^4 D} \sum_{n,s=1}^{m_f} \frac{P_{ns}}{(n^2/b^2 + s^2/a^2)^2} \sum_{i=1}^{N_\eta} \sin(n\pi x_i/b) u_{is}$$

$$- \frac{1}{2\pi^4 D} \sum_{n,s=1}^{m_f} \frac{P_{ns}}{(n^2/b^2 + s^2/a^2)^2} \sum_{i=N_\eta+1}^{N_\eta+N_\xi} \sin(s\pi x_i/a) y_{in} \underset{\substack{(3..3), \\ (2.1.10)}}{=} \frac{1}{2\pi^4 D} \sum_{n,s=1}^{m_f} \frac{abP_{ns}^2}{4(n^2/b^2 + s^2/a^2)^2}$$

$$- \frac{1}{2\pi^4 D} \frac{1}{2a^2} \sum_{i=1}^{N_\eta} \varepsilon_i \sum_{s=1}^{m_f} s^4 \left( \sum_{n=1}^{m_f} \frac{P_{ns} \sin(n\pi x_i/b)}{(n^2/b^2 + s^2/a^2)^2} \right)^2 - \frac{1}{2\pi^4 D} \frac{b}{2a^3} \sum_{j=N_\eta+1}^{N_\eta+N_\xi} \varepsilon_j \sum_{n=1}^{m_f} n^4 \left( \sum_{s=1}^{m_f} \frac{P_{ns} \sin(s\pi x_j/a)}{(n^2/b^2 + s^2/a^2)^2} \right)^2 \quad (5.1.6)$$

This asymptotic expansion obtained for the energy, $U$, consists of a constant term, $U_0$, from which two other terms are subtracted. Naturally then the energy, $U$, attains a minimum if and only if the terms that are subtracted from the constant, $U_0$, attain maxima with respect to $\mathbf{x}$. As the domain is unconstrained, obviously, the subtracted *sums of squares* attain their maxima when each of the *squares* in the sums attains its maximum. As the first sum in the rightmost part of (5.1.6) is symmetric with respect to $i$ and the second sum is symmetric with respect to $j$, clearly:

$$x_i = x_\eta \quad \forall \quad 1 \le i \le N_\eta \quad ; \quad x_j = x_\xi \quad \forall \quad N_\eta + 1 \le j \le N_\eta + N_\xi \quad (5.1.7)$$

is *a globally optimal* solution.

In other words, the energy of a bidirectionally stiffened plate attains a global minimum with respect to the locations of two orthogonal sets of straight, parallel, uniform, infinitesimally rigid stiffeners, for the case of a *general* lateral loading distribution, if all of the stiffeners in *each* of the two sets are located at the *same place*. Knowing the optimal layout one can calculate the energy at the global optimum, substituting (5.1.7) into (5.1.6):

$$U_{\min}(N_\xi, N_\eta) \xrightarrow[\min_i \varepsilon_i \ll 1]{} \frac{1}{2\pi^4 D} \sum_{n,s=1}^{m_f} \frac{abP_{ns}^2}{4(n^2/b^2 + s^2/a^2)^2}$$

$$-\frac{1}{2\pi^4 D} \frac{1}{2a^2} \sum_{s=1}^{m_f} s^4 \left( \sum_{n=1}^{m_f} \frac{P_{ns}\sin(n\pi x_\xi/b)}{(n^2/b^2 + s^2/a^2)^2} \right)^2 \sum_{i=1}^{N_\eta} \varepsilon_i - \frac{1}{2\pi^4 D} \frac{b}{2a^3} \sum_{n=1}^{m_f} n^4 \left( \sum_{s=1}^{m_f} \frac{P_{ns}\sin(s\pi x_\eta/a)}{(n^2/b^2 + s^2/a^2)^2} \right)^2 \sum_{j=N_\eta+1}^{N_\eta+N_\xi} \varepsilon_j \tag{5.1.8}$$

The same energy corresponds to a plate stiffened with one "horizontal" stiffener and one "vertical" stiffener having, respectively, the following bending rigidities:

$$E_\eta I_\eta \equiv aD\sum_{i=1}^{N_\eta} \varepsilon_i \; ; \; E_\xi I_\xi \equiv aD \sum_{j=N_\eta+1}^{N_\eta+N_\xi} \varepsilon_j \tag{5.1.9}$$

Therefore if one has a certain amount of material in the form of two orthogonal sets of parallel highly flexible stiffeners that can be integrally attached to a given plate to minimize the work done by external loads, a globally optimal way to use this material is to split it between exactly two orthogonal beams.

For the special case of a symmetric loading, $P_{ns}=P_{sn}$, and a square plate, $a = b$, the result given in (5.1.8) shows that for flexible enough stiffeners, stiffening with two orthogonal beams is energetically equivalent to stiffening with a single beam, having twice the *bending rigidity*.

This analytically obtained result coincides with numerical observation concerning the same matter, given in the work of Dems et al. (1989).

The power of this result is that a globally optimal stiffening layout is determined analytically for bidirectional stiffening for an arbitrary number of stiffeners having arbitrary cross-sectional parameters and for a general loading case. The limitation of this result is that it is only relevant for highly flexible stiffeners and that it does not say for what actual bending rigidities it is true, for what relative bending rigidities does it cease to be true, and what happens to the optimal positions of the stiffeners for stiffeners having higher bending rigidities.

The reason for this shortcoming is the fact that first order asymptotics with respect to the stiffeners do not capture the 'split' of the stiffeners as an optimal stiffening topology. The relative stiffness for which the first order approximation is valid is not high enough for it to be optimal for the stiffeners to be positioned separately.

The following section shows that, at least for the special case of a bisinusoidal loading and unidirectional stiffening, with two beams positioned symmetrically with respect to the mid-span of the plate, even for a *second order* asymptotic expansion of the energy of the stiffened plate with respect to the relative stiffness of the beams, the degenerated layout given in (5.1.7) is still globally optimal.

A comparison between the first and the second order expansions of the energy enables one to determine for what actual rigidities it is still optimal to place the stiffeners together.

## 5.2 Highly flexible unidirectional stiffeners with no torsional rigidity – 2$^{nd}$ order asymptotics

Repeating the procedure shown in the preceding subsection and retaining terms up to the order of $O(\varepsilon^2)$,

$$\mathbf{g}_n \underset{(3.7)}{=} (\mathbf{1}+\underbrace{\mathbf{F}_n}_{O(\varepsilon)})^{-1}\left[\mathbf{y}_n - \sum_{s=1}^\infty \mathbf{D}_{ns}(\mathbf{1}+\underbrace{\mathbf{B}_s}_{O(\varepsilon)})^{-1}\mathbf{u}_s\right] \xrightarrow[\varepsilon\ll 1]{} \underbrace{\mathbf{y}_n}_{O(\varepsilon)} - \underbrace{\mathbf{F}_n}_{O(\varepsilon)}\underbrace{\mathbf{y}_n}_{O(\varepsilon)} - \sum_{s=1}^\infty \underbrace{\mathbf{D}_{ns}}_{O(\varepsilon)}\underbrace{\mathbf{u}_s}_{O(\varepsilon)} \tag{5.2.1}$$

$$\mathbf{M}_{nm} \underset{(3.7)}{=} \sum_{s=1}^\infty (\mathbf{1}+\underbrace{\mathbf{F}_n}_{O(\varepsilon)})^{-1}\mathbf{D}_{ns}(\mathbf{1}+\underbrace{\mathbf{B}_s}_{O(\varepsilon)})^{-1}\mathbf{C}_{sm} \xrightarrow[\varepsilon\ll 1]{} \sum_{s=1}^\infty \underbrace{\mathbf{D}_{ns}}_{O(\varepsilon)}\underbrace{\mathbf{C}_{sm}}_{O(\varepsilon)}$$

$$\Rightarrow \bar{\mathbf{z}} \underset{(3.9)}{=} (\mathbf{1}-\underbrace{\mathbf{M}}_{O(\varepsilon^2)})^{-1}\bar{\mathbf{g}} \xrightarrow[\varepsilon\ll 1]{} (\mathbf{1}+\underbrace{\mathbf{M}}_{O(\varepsilon^2)})\bar{\mathbf{g}} \xrightarrow[\varepsilon\ll 1]{} \underbrace{\bar{\mathbf{g}}}_{O(\varepsilon)} + \underbrace{\overline{\mathbf{Mg}}}_{O(\varepsilon^3)} \to \underbrace{\bar{\mathbf{g}}}_{O(\varepsilon)} \Rightarrow \mathbf{z}_n \xrightarrow[\varepsilon\ll 1]{} \mathbf{y}_n - \mathbf{F}_n\mathbf{y}_n - \sum_{s=1}^\infty \mathbf{D}_{ns}\mathbf{u}_s \tag{5.2.2}$$

$$\mathbf{v}_r \underset{(3.13)}{=} (\mathbf{1}+\underbrace{\mathbf{B}_r}_{O(\varepsilon)})^{-1}\left(\mathbf{u}_r - \sum_{n=1}^{m_f}\mathbf{C}_{rn}\mathbf{z}_n\right) \xrightarrow[\varepsilon\ll 1]{} \mathbf{u}_r - \mathbf{B}_r\mathbf{u}_r - \sum_{n=1}^{m_f}\mathbf{C}_{rn}\mathbf{z}_n \xrightarrow[\varepsilon\ll 1]{} \mathbf{u}_r - \mathbf{B}_r\mathbf{u}_r - \sum_{n=1}^\infty \mathbf{C}_{rn}\mathbf{y}_n \tag{5.2.3}$$

and substituting them into the expression for the work done by the external loads given in (5.1.5), one gets:

$$\Rightarrow U(\mathbf{x}, N_\eta, N_\xi) \underset{\varepsilon \ll 1}{\to} \frac{1}{2\pi^4 D} \sum_{n,s=1}^{m_f} \frac{P_{ns}}{(n^2/b^2 + s^2/a^2)^2} \left[ \frac{abP_{ns}}{4} - \sum_{i=1}^{N_\eta} \sin(n\pi x_i/b) u_{is} - \sum_{i=N_\eta+1}^{N_\eta+N_\xi} \sin(s\pi x_i/a) y_{in} \right] +$$

$$+ \frac{1}{2\pi^4 D} \sum_{n,s=1}^{m_f} \frac{P_{ns}}{(n^2/b^2 + s^2/a^2)^2} \left[ \sum_{i,j=1}^{N_\eta} \sin(n\pi x_i/b) B_{ijs} u_{js} + \sum_{i,j=N_\eta+1}^{N_\eta+N_\xi} \sin(s\pi x_i/a) F_{ijn} y_{jn} \right] + \quad (5.2.4)$$

$$+ \frac{1}{2\pi^4 D} \sum_{n,s=1}^{m_f} \frac{P_{ns}}{(n^2/b^2 + s^2/a^2)^2} \left[ \sum_{r=1}^{m_f} \sum_{i=1}^{N_\eta} \sum_{j=N_\eta+1}^{N_\eta+N_\xi} \sin(n\pi x_i/b) C_{ijsr} y_{jr} + \sum_{k=1}^{m_f} \sum_{i=N_\eta+1}^{N_\eta+N_\xi} \sum_{j=1}^{N_\eta} \sin(s\pi x_i/a) D_{ijnk} u_{jk} \right]$$

Next, recalling (3.6) and rearranging, results in the following 2$^{\text{nd}}$ order asymptotic expansion of $U$:

$$\Rightarrow U(\mathbf{x}, N_\xi, N_\eta) \underset{\varepsilon \ll 1}{\to} \frac{ab}{8\pi^4 D} \sum_{n,s=1}^{\infty} \frac{P_{ns}^2}{(n^2/b^2 + s^2/a^2)^2}$$

$$- \frac{1}{4\pi^4 a^3 D^2} \sum_{i=1}^{N_\eta} E_i I_i \sum_{s=1}^{\infty} \left[ \sum_{n=1}^{\infty} \frac{s^2 \sin(n\pi x_i/b) P_{ns}}{(n^2/b^2 + s^2/a^2)^2} \right]^2 - \frac{b}{4\pi^4 a^4 D^2} \sum_{i=N_\eta+1}^{N_\eta+N_\xi} E_i I_i \sum_{n=1}^{\infty} \left[ \sum_{s=1}^{\infty} \frac{n^2 \sin(s\pi x_i/a) P_{ns}}{(n^2/b^2 + s^2/a^2)^2} \right]^2 +$$

$$+ \frac{1}{2\pi^4 a^7 b D^3} \sum_{i,j=1}^{N_\eta} \sum_{s=1}^{\infty} \sum_{\mu=1}^{\infty} \frac{\sin(\mu\pi x_i/b) \sin(\mu\pi x_j/b)}{(\mu^2/b^2 + s^2/a^2)^2} \sum_{n=1}^{\infty} \frac{E_i I_i s^4 \sin(n\pi x_i/b) P_{ns}}{(n^2/b^2 + s^2/a^2)^2} \sum_{m=1}^{\infty} \frac{E_j I_j s^4 \sin(m\pi x_j/b) P_{ms}}{(m^2/b^2 + s^2/a^2)^2} + \quad (5.2.5)$$

$$+ \frac{b}{2\pi^4 a^9 D^3} \sum_{i,j=N_\eta+1}^{N_\eta+N_\xi} \sum_{n=1}^{\infty} \sum_{w=1}^{\infty} \frac{\sin(w\pi x_i/a) \sin(w\pi x_j/a)}{(n^2/b^2 + w^2/a^2)^2} \sum_{s=1}^{\infty} \frac{E_i I_i n^4 \sin(s\pi x_i/a) P_{ns}}{(n^2/b^2 + s^2/a^2)^2} \sum_{q=1}^{\infty} \frac{E_j I_j n^4 \sin(q\pi x_j/a) P_{nq}}{(n^2/b^2 + q^2/a^2)^2} +$$

$$+ \frac{1}{2\pi^4 a^8 D^3} \sum_{i=1}^{N_\eta} \sum_{j=N_\eta+1}^{N_\eta+N_\xi} \sum_{r,s=1}^{\infty} \frac{\sin(r\pi x_i/b) \sin(s\pi x_j/a)}{(r^2/b^2 + s^2/a^2)^2} \sum_{n=1}^{\infty} \frac{E_i I_i s^4 \sin(n\pi x_i/b) P_{ns}}{(n^2/b^2 + s^2/a^2)^2} \sum_{q=1}^{\infty} \frac{E_j I_j r^4 \sin(q\pi x_j/a) P_{rq}}{(r^2/b^2 + q^2/a^2)^2} +$$

$$+ \frac{1}{2\pi^4 a^8 D^3} \sum_{j=1}^{N_\eta} \sum_{i=N_\eta+1}^{N_\eta+N_\xi} \sum_{n,k=1}^{\infty} \frac{\sin(n\pi x_j/b) \sin(k\pi x_i/a)}{(n^2/b^2 + k^2/a^2)^2} \sum_{m=1}^{\infty} \frac{E_j I_j k^4 \sin(m\pi x_j/b) P_{mk}}{(m^2/b^2 + k^2/a^2)^2} \sum_{s=1}^{\infty} \frac{E_i I_i n^4 \sin(s\pi x_i/a) P_{ns}}{(n^2/b^2 + s^2/a^2)^2}$$

At this point, as the result is anyway an asymptotic one, several assumptions are made in order to get an analytic form of the solution of the optimal stiffening problem. First, the loading is assumed to be bisinusoidal, with half-periods equal to the lengths of the sides of the plate. Second, for simplicity and clarity of illustration, unidirectional stiffening is assumed, with only two stiffeners, positioned symmetrically with respect to the center of the plate (the a priori symmetry of the stiffening layout seems justified in view of the symmetry of the loading). With these assumptions, the work done by the loading becomes:

$$U(\delta, \varepsilon, p_0, a, b, D) \underset{\varepsilon \ll 1}{\to} \frac{abp_0^2}{8\pi^4 D} \frac{1}{(1/b^2 + 1/a^2)^2} - \frac{p_0^2 \varepsilon}{4\pi^4 a^2 D} \frac{\{\sin^2(\pi[1/2 + \delta]) + \sin^2(\pi[1/2 - \delta])\}}{(1/b^2 + 1/a^2)^4} +$$

$$+ \frac{p_0^2 \varepsilon^2}{2\pi^4 a^5 bD} \frac{\sin^2(\pi[1/2 + \delta])}{(1/b^2 + 1/a^2)^4} \sum_{n=1}^{\infty} \frac{\sin^2(n\pi[1/2 + \delta])}{(n^2/b^2 + 1/a^2)^2} +$$

$$+ \frac{p_0^2 \varepsilon^2}{\pi^4 a^5 bD} \frac{\sin(\pi[1/2 + \delta]) \sin(\pi[1/2 - \delta])}{(1/b^2 + 1/a^2)^4} \sum_{n=1}^{\infty} \frac{\sin(n\pi[1/2 + \delta]) \sin(n\pi[1/2 - \delta])}{(n^2/b^2 + 1/a^2)^2} + \quad (5.2.6)$$

$$+ \frac{p_0^2 \varepsilon^2}{2\pi^4 a^5 bD} \frac{\sin^2(\pi[1/2 - \delta])}{(1/b^2 + 1/a^2)^4} \sum_{n=1}^{\infty} \frac{\sin^2(n\pi[1/2 - \delta])}{(n^2/b^2 + 1/a^2)^2}$$

where $\delta$ is the distance of each of the stiffeners from the center of the plate divided by $b$, $\varepsilon \equiv EI/aD$ and $a$ is the length of each of the stiffeners. Employing trigonometric identities and recalling the definition: $\beta \equiv b/a$ one can obtain a simpler expression for the (non-dimensionalized) elastic strain energy of the loaded structure, consistent of the plate and the stiffeners:

$$\hat{U}(\delta, \varepsilon, \beta) \underset{\varepsilon \ll 1}{\to} \frac{\beta^5}{(\beta^2+1)^2} \left\{ 1 - (4\beta^3 \varepsilon) \frac{\cos^2(\pi\delta)}{(\beta^2+1)^2} \left[ 1 + 2\beta^3 \varepsilon \sum_{n=1,3,5}^{\infty} \frac{1}{(\beta^2+n^2)^2} + 2\beta^3 \varepsilon \sum_{n=1,3,5}^{\infty} \frac{\cos(2n\pi\delta)}{(\beta^2+n^2)^2} \right] \right\} \quad (5.2.7)$$

where the actual energy can be obtained from the non-dimensional form by:

$$U(\delta, \varepsilon, \beta, a, p_0, D) = \frac{a^6 p_0^2}{8\pi^4 D} \hat{U}(\delta, \varepsilon, \beta) \quad (5.2.8)$$

Next, the series in (5.2.7) can be expressed in terms of the Fourier cosine series of cosh$z$ and $z$sinh$z$, $z$ being a linear function of $|\delta|$. Thence, substituting these functions instead of their Fourier series representations, one obtains:

$$\hat{U}(\delta,\beta,\varepsilon) \underset{\varepsilon \ll 1}{\to} \frac{\beta^5}{(\beta^2+1)^2} - \frac{4\beta^8 \varepsilon \cos^2(\pi\delta)}{(\beta^2+1)^4}$$

$$- \frac{8\beta^{11}\varepsilon^2 \cos^2(\pi\delta)}{(\beta^2+1)^4}\left[\frac{\pi\coth(\pi\beta)}{4\beta^3} + \frac{\pi^2 \coth^2(\pi\beta)}{4\beta^2} - \frac{\pi\coth(\pi\beta/2)}{8\beta^3} - \frac{\pi^2 \coth^2(\pi\beta/2)}{16\beta^2} - \frac{3\pi^2}{16\beta^2}\right] \quad (5.2.9)$$

$$- \frac{8\beta^{11}\varepsilon^2 \cos^2(\pi\delta)}{(\beta^2+1)^4}\left\{\frac{\pi\cosh[\pi\beta(1-2|\delta|)]}{4\beta^3 \sinh(\pi\beta)} + \frac{\pi^2 \cosh(\pi\beta)\cosh[\pi\beta(1-2|\delta|)]}{4\beta^2 \sinh^2(\pi\beta)} - \frac{\pi^2(1-2|\delta|)\sinh[\pi\beta(1-2|\delta|)]}{4\beta^2 \sinh(\pi\beta)}\right.$$

$$\left.- \frac{\pi\cosh[\pi\beta(1/2-2|\delta|)]}{8\beta^3 \sinh(\pi\beta/2)} - \frac{\pi^2 \cosh(\pi\beta/2)\cosh[\pi\beta(1/2-2|\delta|)]}{16\beta^2 \sinh^2(\pi\beta/2)} + \frac{\pi^2(1-4|\delta|)\sinh[\pi\beta(1/2-2|\delta|)]}{16\beta^2 \sinh(\pi\beta/2)}\right\}$$

Differentiating with respect to $\delta$ one gets:

$$\frac{\partial}{\partial \delta}\hat{U}(\delta,\beta,\varepsilon) \underset{\varepsilon \ll 1}{\to} \frac{4\pi\beta^8 \varepsilon \sin(2\pi\delta)}{(\beta^2+1)^4} +$$

$$+ [4\coth(\pi\beta) + 4\pi\beta\coth^2(\pi\beta) - 2\coth(\pi\beta/2) - \pi\beta\coth^2(\pi\beta/2) - 3\pi\beta]\frac{\pi^2 \beta^8 \varepsilon^2 \sin(2\pi\delta)}{2(\beta^2+1)^4} +$$

$$+ \frac{\pi^2 \beta^8 \varepsilon^2 \sin(2\pi\delta)}{2(\beta^2+1)^4}\left\{\frac{4[\sinh(\pi\beta) + \pi\beta\cosh(\pi\beta)]\cosh[\pi\beta(1-2|\delta|)]}{\sinh^2(\pi\beta)} - \frac{4\pi\beta(1-2|\delta|)\sinh[\pi\beta(1-2|\delta|)]}{\sinh(\pi\beta)}\right. \quad (5.2.10)$$

$$\left.- \frac{[2\sinh(\pi\beta/2) + \pi\beta\cosh(\pi\beta/2)]\cosh[\pi\beta(1/2-2|\delta|)]}{\sinh^2(\pi\beta/2)} + \frac{\pi\beta(1-4|\delta|)\sinh[\pi\beta(1/2-2|\delta|)]}{\sinh(\pi\beta/2)}\right\}$$

$$- \frac{\beta^{10}\varepsilon^2 \cos^2(\pi\delta)}{(\beta^2+1)^4}\left\{\frac{4\pi^3(1-2|\delta|)\cosh[\pi\beta(1-2|\delta|)]}{\sinh(\pi\beta)} - \frac{4\pi^3 \cosh(\pi\beta)\sinh[\pi\beta(1-2|\delta|)]}{\sinh^2(\pi\beta)} + \right.$$

$$\left.+ \frac{\pi^3 \cosh(\pi\beta/2)\sinh[\pi\beta(1/2-2|\delta|)]}{\sinh^2(\pi\beta/2)} - \frac{\pi^3(1-4|\delta|)\cosh[\pi\beta(1/2-2|\delta|)]}{\sinh(\pi\beta/2)}\right\}\text{sgn}(\delta)$$

Now, in order to determine whether placing the stiffeners together is optimal, one could have examined whether the second derivative of the energy with respect to $\delta$ at $\delta=0$ changes its sign as $\varepsilon$ increases. Such a change of sign would have indicated that unified stiffeners cease to correspond to a global minimum of the energy and start to correspond to its local maximum as their rigidity is increased. This would imply that a second order asymptotic expansion of the energy enables one to find a critical relative rigidity, $\varepsilon$, above which the stiffeners are best to be split and that this critical relative rigidity is small enough for the second order approximation to be a good one.

However, the second derivative with respect to $\delta$ of the second order expansion of the energy with respect to $\varepsilon$ is undefined at $\delta=0$ due to the non-continuousness of the sign function.

Therefore, instead, one can check whether the first derivative vanishes anywhere besides at $\delta=0$ as $\varepsilon$ increases. In fact, it is enough to check the sign of $\lim_{\delta \to 0^+} \delta^{-1} \partial \hat{U}/\partial \delta$. If the sign is positive for every positive $\varepsilon$, then placing the stiffeners together never corresponds to a local maximum of the second order expansion of the energy.

Section 5.1 shows that placing the stiffeners together is globally optimal in the limit of a small positive $\varepsilon$. Therefore if $\delta=0$ happens to be a *local* minimum of the energy for every $\varepsilon$ in the range of the validity of the second order asymptotics and there is another, global, minimum elsewhere then, due to the continuity of the energy with respect to both $\delta$ and $\varepsilon$ there has to be a local maximum between these two local minima. Numerical examination shows that when $\varepsilon$ is high enough such that $\delta=0$ ceases to be the global optimum, the globally optimal solution corresponds to $\delta$ continuously increasing with $\varepsilon$. Therefore in the limit case where $\varepsilon$ is only infinitesimally higher than its critical value, the energy would have three extrema (two local minima and a local maximum in between) located at a line segment with a length tending to zero. This would imply infinite derivatives and therefore contradict with the fact that the energy is Hölder continuous in the vicinity of zero (as one can show by expanding $\delta^{-1}\partial\hat{U}/\partial\delta$ as a Taylor series around $\delta=0$. By expanding $\delta^{-1}\partial \hat{U}/\partial \delta$ rather than $\partial \hat{U}/\partial \delta$ one eliminates the non-continuous functions). This contradiction can only be avoided by the assumption that wherever $\delta = 0$ is a local minimum of the energy, it is also a global minimum and that in order for it not to be a global minimum it has to be a local maximum. Also, a function Hölder continuous in the vicinity of zero behaves as a power function near zero and thus cannot have three extrema within a line segment starting at zero and having an arbitrarily small length.

Hence, by proving that $\lim_{\delta \to 0^+} \delta^{-1}\partial \hat{U}/\partial \delta$ is positive for every $\varepsilon$ for which the second order expansion of the energy is valid, one proves that placing the stiffeners together is globally optimal (within the abovementioned range of rigidities).

Next, expanding the hyperbolic functions of $\delta$ in (5.2.10) using the identities of hyperbolic trigonometry, dividing by $\delta$ and computing the limits, recalling that:

$$\sinh(2\pi\beta|\delta|)\text{sgn}(\delta)/[2\pi\beta\delta] = \sinh(2\pi\beta\delta)/[2\pi\beta\delta] \underset{\delta \to 0}{\to} 1 \tag{5.2.11}$$

the following result is obtained:

$$\lim_{\delta \to 0^+} \delta^{-1}\partial \hat{U}/\partial \delta \underset{\varepsilon \ll 1}{\to} \frac{8\pi^2 \beta^8 \varepsilon(1-\varepsilon/\varepsilon_{cr})}{(\beta^2+1)^4} \tag{5.2.12}$$

where

$$\varepsilon_{cr} \equiv \frac{4}{\pi\{2(1+\beta^2)[\coth(\pi\beta/2) - 2\coth(\pi\beta)] + \pi\beta(\beta^2-1)[4\coth^2(\pi\beta) - \coth^2(\pi\beta/2) - 3]\}} \tag{5.2.13}$$

Obviously, the limit in (5.2.13) is positive if $\varepsilon_{cr}$ is negative for every positive $\beta$. Asymptotic expansions of $\varepsilon_{cr}(\beta)$ show that this is indeed the case for the limit cases of $\beta \ll 1$, $\beta \gg 1$ and $|1-\beta| \ll 1$:

$$\varepsilon_{cr} \underset{\beta \gg 1}{\to} -\frac{2}{\pi(1+\beta^2)} < 0, \quad \varepsilon_{cr} \underset{\beta \ll 1}{\to} -\frac{24}{\pi^2\beta^3[12+5\pi^2\beta(1-\beta)]} < 0 \tag{5.2.14}$$

$$\varepsilon_{cr} \underset{|1-\beta| \ll 1}{\to} -\pi^{-1}[2\coth(\pi) - \coth(\pi/2)]^{-1} \left\{ 1 + \left[ 1 + \pi \frac{3 + \coth^2(\pi/2) - 4\coth^2(\pi)}{2\coth(\pi) - \coth(\pi/2)} \right](1-\beta) \right\} \cong \tag{5.2.15}$$

$$\cong -0.8831(1 - 0.6069\beta) < 0 \quad (\forall \beta < 1.6477)$$

Numerical verification shows that $\varepsilon_{cr}(\beta)$ is negative for all positive finite values of $\beta$. Therefore the limit in (5.2.12) is positive and thus for small enough positive *finite* values of $\varepsilon$ it is globally optimal to stiffen a simply supported thin rectangular plate in elastostatic bending under a lateral bisinusoidal loading with two stiffeners positioned *at the same place*, in the geometric center of the plate, in parallel to one of the edges of the plate. The fact that the global optimality of this stiffening layout holds for *finite* values of $\varepsilon$, at least for the specified loading, is due to the fact that this result is true for both the first and the second order asymptotic expansions of the elastic energy of the stiffened plate, and it is the essence of the present subsection. The (non-dimensionalized) structural strain energy corresponding to the globally optimal layout discussed above can be obtained in a second order approximation with respect to $\varepsilon$ by setting $\delta = 0$ in (5.2.9):

$$\hat{U}^{opt}(\beta, \varepsilon) \underset{\varepsilon \ll 1}{\to} \frac{\beta^5}{(\beta^2+1)^2} - \frac{4\beta^8 \varepsilon}{(\beta^2+1)^4}[1 + f(\beta)\varepsilon] \tag{5.2.16}$$

where:

$$f(\beta) \equiv [4\coth(\pi\beta) + 4\pi\beta\coth^2(\pi\beta) - 2\coth(\pi\beta/2) - \pi\beta\coth^2(\pi\beta/2) - 3\pi\beta]\pi/4 \tag{5.2.17}$$

Both the first and the second order approximations of the energy hold for small enough values of $\varepsilon$, formally, when $f(\beta)\varepsilon = \gamma = o(1)$. Then the higher orders contribute negligibly to the first and second order approximations of the energy for which the degenerated stiffening layout is proven herein to be globally optimal for general and bisinusoidal loadings, respectively. Recalling the definition of the relative bending rigidity and assuming blade stiffeners with cross-sectional width $d$ and static height $h$ and a plate with thickness $t$ and Poisson's ratio $\nu$, and that the plate and the stiffener have the same Young modulus, one can obtain an approximation for the maximal static height of the stiffener for which it is still globally optimal not to split it. Hence, assuming that the cross-sectional width of the stiffener is much smaller than its length, as one might expect from a beam, namely that $d/a = \zeta = o(1)$, and making the approximation $\gamma \sim \zeta$ based on the fact that both these quantities should be smaller than unity and are compliant to no additional restrictions (the values of $\gamma$ and $\zeta$ can each be, say, 0.01), and, last, taking $\nu$ to be 1/3, which is typical for structural materials, one obtains the following approximation:

$$\frac{h}{t} < \frac{1}{2}\left[\frac{9}{f(\beta)}\right]^{1/3} \qquad (5.2.18)$$

Taking the limit of (5.2.17) for $\beta \gg 1$ one obtains the result that for a short stiffener the critical static height is $h_{max} \sim [9/(4\pi)]^{1/3} t \cong 0.9t$. For a square plate, $\beta = 1$, $h_{max} \sim t$. That is, if the stiffener is not relatively long, it can have about the same static height as the thickness of the plate and it would still be flexible enough for it to be globally optimal not to split it. This result is impractical, since the stiffener is assumed to be placed symmetrically with respect to the mid-surface of the plate and thus if its static height is equal to the thickness of the plate and Young's modulus of the plate and the stiffener are the same, then it is as having no stiffener at all, and it does not matter whether it is split or not. However, if the stiffener is longer, its length makes it more flexible such that its static height can be larger for the optimality of stiffening with a single stiffener yet to hold. For example, it follows from (5.2.17) and (5.2.18) that for the case $\beta = 1/2$ the critical static height is $h_{max} \sim 1.5t$, for $\beta = 1/3$, $h_{max} \sim 2t$ and for $\beta = 1/6$, $h_{max} \sim 4t$.

This simple calculation shows that, at least for long stiffeners, the degenerated stiffening layout is globally optimal for rather realistic parameters and not only, as one could presume, for stiffeners of absolutely negligible bending rigidity.

In the results obtained above, the globally optimal layout corresponds to a thin plate with thin stiffeners. The fact that the bending rigidity of the stiffeners normalized by the bending rigidity of the plate is small does not imply that for the stiffeners to be realistic the plate should be thick, since a stiffener of width and height of the order of the thickness of the plate has a much smaller bending rigidity than that of the plate, as it appears from the definition of the relative rigidity, $\varepsilon \equiv EI/aD$, which is proportional to the ratio of the width of the stiffener and the breadth of the plate, which is usually a small number. Thus the above derivations do not contradict with the assumptions of the first order Love-Kirchhoff plate bending theory.

The following subsection examines the opposite case, where the stiffeners are highly rigid compared to the plate. If flexible stiffeners have an asymptotically optimal layout in which they are combined to form a single stiffener, then, as is it turns out, highly rigid stiffeners have their own asymptotically optimal layout, where they are optimally spread, and to which they tend as their rigidities increase.

### 5.3 Highly rigid unidirectional stiffeners with no torsional rigidity – 1st order asymptotics

The present subsection examines the case where the stiffeners are highly rigid with respect to the plate. By means of a first order asymptotic analysis with respect to the *flexibilities* of the stiffeners, it is shown that rigid enough stiffeners are best to be positioned separately, in locations determined by the functionality of the loading. It is shown that the optimal positions of the stiffeners approach their asymptotic limits as the rigidities of the stiffeners approach infinity. The functionality of this approach is given analytically for bisinusoidal loading with half-periods equal to the lengths of the edges of the plate, unidirectional stiffening with two stiffeners positioned symmetrically with respect to the mid-span of the plate and the condition that the stiffeners are relatively long. Substituting the expression for the deflection field for the case, as given in (2.20) into (2.12) and making use of (2.9), employing the definitions of $\beta$ and $\varepsilon$ used in the previous subsections, and defining $\hat{x} \equiv x/b$ one obtains the strain energy of the stiffened plate, as follows:

$$U(\hat{x}) = \frac{\beta P^2 a^6}{8\pi^4 D(\beta^{-2}+1)^2} - \frac{P^2 a^6 \sin(\pi\hat{\mathbf{x}}^T)\left(\varepsilon^{-1}\mathbf{1}+\hat{\mathbf{B}}\right)^{-1}\sin(\pi\hat{\mathbf{x}})}{4\pi^4 D(\beta^{-2}+1)^4} \qquad (5.3.1)$$

where $\hat{\mathbf{x}} = (\hat{x}, 1-\hat{x})^T$.

Evaluating expression (2.17) for the abovementioned assumptions and dividing it by $\varepsilon$, the coefficients defining the matrix in (5.3.1) can be readily obtained:

$$\hat{B}_{11} = \frac{1}{2\pi^3}\left\{\beta\frac{1-\cosh(2\pi\beta\hat{x})}{\sinh^2(\pi\beta)} + \frac{\cosh(\pi\beta)-\cosh(\pi\beta[1-2\hat{x}])}{\pi\sinh(\pi\beta)} - 2\beta\frac{\hat{x}\sinh(\pi\beta[1-2\hat{x}])}{\sinh(\pi\beta)}\right\}$$

$$\hat{B}_{22} = \frac{1}{2\pi^3}\left\{\beta\frac{1-\cosh(2\pi\beta[1-\hat{x}])}{\sinh^2(\pi\beta)} + \frac{\cosh(\pi\beta)-\cosh(\pi\beta[1-2\hat{x}])}{\pi\sinh(\pi\beta)} + 2\beta\frac{[1-\hat{x}]\sinh(\pi\beta[1-2\hat{x}])}{\sinh(\pi\beta)}\right\} \qquad (5.3.2)$$

$$\hat{B}_{12} = \hat{B}_{21} = \frac{1}{2\pi^3}\left\{\beta\frac{\cosh(\pi\beta[1-2\hat{x}])-\cosh(\pi\beta)}{\sinh^2(\pi\beta)} + \frac{\cosh(2\pi\beta\hat{x})-1}{\pi\sinh(\pi\beta)} + \beta\frac{(1-2\hat{x})\sinh(2\pi\beta\hat{x})}{\sinh(\pi\beta)}\right\}$$

Considering $\varepsilon$ large enough for the spectral norm of $\varepsilon^{-1}\hat{\mathbf{B}}^{-1}$ to be much smaller than unity and expanding the expression for the energy in (5.3.1) into a first order Taylor series of the relative flexibility of the stiffeners, $\varepsilon^{-1}$, one obtains the following asymptotic relation:

$$U(\hat{x}) \underset{\varepsilon \gg 1}{\rightarrow} \frac{\beta P^2 a^6}{8\pi^4 D(\beta^{-2}+1)^2} - \frac{P^2 a^6 [\sin(\pi\hat{\mathbf{x}}^T)\hat{\mathbf{B}}^{-1}\sin(\pi\hat{\mathbf{x}}) - \varepsilon^{-1}\sin(\pi\hat{\mathbf{x}}^T)\hat{\mathbf{B}}^{-2}\sin(\pi\hat{\mathbf{x}})]}{4\pi^4 D(\beta^{-2}+1)^4} \quad (5.3.3)$$

Noting the symmetry of the sine function, inverting the 2x2 matrices in (5.3.3) analytically, and introducing the auxiliary variables,

$$\overline{B}_{ij} \equiv 2\pi^4 \sinh^2(\pi\beta) \hat{B}_{ij} \quad (5.3.4)$$

one gets the following representation of the energy:

$$U(\hat{x}) \underset{\varepsilon \gg 1}{\rightarrow} \frac{\beta P^2 a^6}{8\pi^4 D(\beta^{-2}+1)^2} - \frac{P^2 a^6 \sinh^2(\pi\beta)}{2D(\beta^{-2}+1)^4} \left[ \frac{\overline{B}_{22} - 2\overline{B}_{12} + \overline{B}_{11}}{\overline{B}_{11}\overline{B}_{22} - \overline{B}_{12}^2} - \varepsilon^{-1} \frac{(\overline{B}_{22} - \overline{B}_{12})^2 + (\overline{B}_{11} - \overline{B}_{12})^2}{(\overline{B}_{11}\overline{B}_{22} - \overline{B}_{12}^2)^2} \right] \sin^2(\pi\hat{x}) \quad (5.3.5)$$

Taking a Taylor series expansion with respect to $\beta$ of the $\overline{B}_{ij}$ components for a small $\beta$ and retaining only the first non-vanishing powers of $\beta$ results in:

$$\overline{B}_{12} = \overline{B}_{21} \underset{\beta \ll 1}{\rightarrow} 2(\pi\beta)^5 \hat{x}^2(1-2\hat{x}^2)/3 \quad , \quad \overline{B}_{22} \underset{\beta \ll 1}{\rightarrow} \overline{B}_{11} \underset{\beta \ll 1}{\rightarrow} 4(\pi\beta)^5 \hat{x}^2(1-\hat{x})^2/3 \quad (5.3.6)$$

Substituting (5.3.6) into (5.3.5), one gets an asymptotic representation of the energy for $\beta \ll 1$ (which, due to the high order of the expansion, is a fair approximation for any $\beta<1$):

$$U(\hat{x}) \underset{\substack{\varepsilon \gg 1 \\ \beta \ll 1}}{\rightarrow} \frac{\beta P^2 a^6}{8\pi^4 D(\beta^{-2}+1)^2} - \frac{P^2 a^6 \sinh^2(\pi\beta)}{2D(\beta^{-2}+1)^4 (\pi\beta)^5} \left[ \frac{3\sin^2(\pi\hat{x})}{3\hat{x}^2 - 4\hat{x}^3} - (\pi\beta)^{-5}\varepsilon^{-1} \frac{9\sin^2(\pi\hat{x})}{2(3\hat{x}^2 - 4\hat{x}^3)^2} \right] \quad (5.3.7)$$

The asymptotic limit for the optimal location of long rigid stiffeners can be obtained by minimization of the limit of (5.3.7) for $\varepsilon \rightarrow \infty, 0 < \beta \ll 1, \beta^5 \varepsilon \rightarrow \infty$ (that is finitely long and infinitely rigid stiffeners):

$$\hat{x}_{opt}^{as} = \underset{\hat{x}\in[0,1/2]}{\arg\min} \lim_{\substack{\varepsilon \rightarrow \infty \\ \beta \ll 1 \\ \beta^5 \varepsilon \rightarrow \infty}} U(\hat{x}) = \underset{\hat{x}\in[0,1/2]}{\arg\min} \left[ 1 - \frac{12\pi \sin^2(\pi\hat{x})}{3\hat{x}^2 - 4\hat{x}^3} \right] \cong 0.3445 \quad (5.3.8)$$

Next, substituting the definition $\delta \equiv \hat{x} - \hat{x}_{opt}^{as}$ into (5.3.7), performing a 2nd order asymptotic expansion with respect to $\delta$ around zero, and differentiating with respect to $\delta$, noting that this derivative should vanish for $\delta, \varepsilon^{-1} \rightarrow 0$, one obtains an asymptotic expansion of $\partial U(\varepsilon^{-1}, \beta, \delta \ll 1)/\partial \delta$ (which in this case becomes a linear function of $\delta$). Finding the single root of the abovementioned derivative produces an expression for the globally optimal location of a stiffener (the one closer to the origin of the coordinate system) in a non-dimensional form, $\hat{x}$, as a function of $\varepsilon$ and $\beta$ for $\beta, \varepsilon^{-1}, \beta^{-5}\varepsilon^{-1} \ll 1$, as follows:

$$\frac{1/2 - \hat{x}_{opt}}{1/2 - \hat{x}_{opt}^{as}} \underset{\varepsilon^{-1}, \beta, \beta^{-5}\varepsilon^{-1} \ll 1}{\rightarrow} 1 - q\beta^{-5}\varepsilon^{-1} \quad (5.3.9)$$

where $q$ is:

$$q = \frac{\dfrac{3}{2\pi^5} \dfrac{1}{(\hat{x}_{opt}^{as})^2} \left[ \dfrac{\cot(\pi\hat{x}_{opt}^{as})}{(1/2 - \hat{x}_{opt}^{as})} - \dfrac{12}{\hat{x}_{opt}^{as}(3 - 4\hat{x}_{opt}^{as})} \right]}{\dfrac{12\hat{x}_{opt}^{as} - 3}{(\hat{x}_{opt}^{as})^2} - 12\dfrac{1/2 - \hat{x}_{opt}^{as}}{\hat{x}_{opt}^{as}} + (3 - 4\hat{x}_{opt}^{as})[\cot^2(\pi\hat{x}_{opt}^{as}) - 1] - 24\cot(\pi\hat{x}_{opt}^{as})\dfrac{1/2 - \hat{x}_{opt}^{as}}{\hat{x}_{opt}^{as}}} \cong 0.268 \quad (5.3.10)$$

As an illustration of the result obtained above, one might consider the following stiffening parameters: $\beta = 0.9$ (which seems acceptable due to the high order of the expansion with respect to $\beta$), $\nu = 1/3$, $d/a = 0.01$ and $h/t = 10$, which imply $\varepsilon = 11.25$. According to (5.3.9) and (5.3.10), the corresponding optimal position is: $\hat{x}_{opt} \cong 0.35$. It is therefore established that the globally optimal stiffening layout for even moderately long stiffeners consists of a single stiffener for a finite range of rigidities, and that for a certain value of the relative rigidity, $\varepsilon$, the stiffener should be split into two. Then, for two *rigid* stiffeners in an optimal layout, the distance between the stiffeners should increase (hyperbolically) with their rigidities until it approaches its asymptotic limit. The rate of this increase would in case of an optimal layout be approximately proportional to the 5th power of the aspect ratio of the plate, defined as the length of the stiffeners divided by the length of the edge of the plate perpendicular to the stiffeners.

## 5.4 Highly flexible unidirectional stiffeners with non-negligible torsional rigidity – 1st order asymptotics

The linear elastostatic strain energy for the case can be derived, using the results of sections 3 and 4, as follows. Introducing the small parameters $\varepsilon \equiv EI/(aD), \gamma \equiv GC/(bD)$ and deriving the auxiliary quantities required for the calculation of the strain energy for the asymptotic case, one gets:

$$\varphi_r \underset{(4.11)}{\equiv} \underbrace{\mathbf{v}_r}_{O(\gamma)} - \sum_{s=1}^{\infty} \underbrace{\Lambda_{rs}}_{O(\gamma)} \underbrace{(1+\mathbf{B}_s)^{-1}}_{O(\varepsilon)} \underbrace{\mathbf{u}_s}_{O(\varepsilon)} \underset{\varepsilon \ll 1}{\to} \underbrace{\mathbf{v}_r}_{O(\gamma)} - \sum_{s=1}^{\infty} \underbrace{\Lambda_{rs}}_{O(\gamma)} \underbrace{(1-\mathbf{B}_s)}_{O(\varepsilon)} \underbrace{\mathbf{u}_s}_{O(\varepsilon)} \underset{\varepsilon \ll 1}{\to} \underbrace{\mathbf{v}_r}_{O(\gamma)}$$

$$\Psi_{rs} \underset{(4.11)}{\equiv} \underbrace{\Omega_{rs}}_{O(\gamma)} - \underbrace{\Lambda_{rs}}_{O(\gamma)} \underbrace{(1+\mathbf{B}_s)^{-1}}_{O(\varepsilon)} \underbrace{\mathbf{T}_s}_{O(\varepsilon)} + \delta_{rs} 1 \underset{\varepsilon \ll 1}{\to} \underbrace{\Omega_{rs}}_{O(\gamma)} - \underbrace{\Lambda_{rs}}_{O(\gamma)} \underbrace{(1-\mathbf{B}_s)}_{O(\varepsilon)} \underbrace{\mathbf{T}_s}_{O(\varepsilon)} + \delta_{rs} 1 \underset{\varepsilon \ll 1}{\to} \delta_{rs} 1 \quad (5.4.1)$$

$$\theta_r \underset{(4.15)}{\equiv} \underbrace{(1+\mathbf{B}_r)^{-1}}_{O(\varepsilon)} \underbrace{\mathbf{u}_r}_{O(\varepsilon)} \underset{\varepsilon \ll 1}{\to} \underbrace{(1-\mathbf{B}_r)}_{O(\varepsilon)} \underbrace{\mathbf{u}_r}_{O(\varepsilon)} \underset{\varepsilon \ll 1}{\to} \underbrace{\mathbf{u}_r}_{O(\varepsilon)} \quad ; \quad \Theta_r \underset{(4.15)}{\equiv} \underbrace{(1+\mathbf{B}_r)^{-1}}_{O(\varepsilon)} \underbrace{\mathbf{T}_r}_{O(\varepsilon)} \underset{\varepsilon \ll 1}{\to} \underbrace{(1-\mathbf{B}_r)}_{O(\varepsilon)} \underbrace{\mathbf{T}_r}_{O(\varepsilon)} \underset{\varepsilon \ll 1}{\to} \underbrace{\mathbf{T}_r}_{O(\varepsilon)}$$

$$\Rightarrow \underbrace{\mathbf{v}_r}_{O(\gamma)} \underset{\gamma \ll 1}{\leftarrow} \varphi_r \underset{(4.12)}{=} \sum_{s=1}^{\infty} \Psi_{rs} \mathbf{t}_s \underset{\gamma \ll 1}{\to} \sum_{s=1}^{\infty} (\delta_{rs} 1) \mathbf{t}_s = \mathbf{t}_r \Rightarrow \mathbf{t}_s \underset{\gamma \ll 1}{\to} \underbrace{\mathbf{v}_s}_{O(\gamma)}$$

$$\Rightarrow \mathbf{v}_r \underset{(4.7)}{=} \underbrace{(1+\mathbf{B}_r)^{-1}}_{O(\varepsilon)} \underbrace{(\mathbf{u}_r}_{O(\varepsilon)} - \underbrace{\mathbf{T}_r}_{O(\varepsilon)} \underbrace{\mathbf{t}_r)}_{O(\varepsilon)} \underset{\varepsilon \ll 1}{\to} \underbrace{(1-\mathbf{B}_r)}_{O(\varepsilon)} \underbrace{(\mathbf{u}_r}_{O(\varepsilon)} - \underbrace{\mathbf{T}_r}_{O(\varepsilon)} \underbrace{\mathbf{t}_r)}_{O(\varepsilon)} \to \underbrace{\mathbf{u}_r}_{O(\varepsilon)} - \underbrace{\mathbf{T}_r}_{O(\varepsilon)} \underbrace{\mathbf{v}_r}_{O(\varepsilon)O(\gamma)} \underset{\varepsilon \ll 1}{\to} \underbrace{\mathbf{u}_r}_{O(\varepsilon)} \quad (5.4.2)$$

Then, substituting (5.4.1) and (5.4.2) into the expression for the deflection field given by (4.17), gives:

$$w(\xi,\eta,\mathbf{x}) \underset{\varepsilon,\gamma \ll 1}{\to} \frac{4}{\pi^4 abD} \sum_{n,s=1}^{\infty} \frac{\sin(n\pi\xi/b)\sin(s\pi\eta/a)}{(n^2/b^2+s^2/a^2)^2} \left[ \frac{ab}{4} P_{ns} - \sin(n\pi\mathbf{x}^T/b)\mathbf{u}_s - \frac{n\pi}{b}\cos(n\pi\mathbf{x}^T/b)\mathbf{v}_s \right] \quad (5.4.3)$$

Integrating it multiplied by half the loading (thus obtaining the expression for the external work or the total structural strain energy),

$$U(\mathbf{x}) \underset{\varepsilon,\gamma \ll 1}{\to} \frac{1}{2\pi^4 D} \sum_{n,s=1}^{\infty} \frac{P_{ns}}{(n^2/b^2+s^2/a^2)^2} \left[ \frac{ab}{4} P_{ns} - \sin(n\pi\mathbf{x}^T/b)\mathbf{u}_s - \frac{n\pi}{b}\cos(n\pi\mathbf{x}^T/b)\mathbf{v}_s \right] \quad (5.4.4)$$

and recalling definitions (2.10, 3.6), (4.6) and (4.10), produces the following asymptotic expression:

$$U(\mathbf{x}) \underset{\varepsilon_i,\gamma_i \ll 1}{\to} \sum_{i=1}^{N} \left\{ \frac{abN^{-1}}{8\pi^4 D} \sum_{n,s=1}^{\infty} \frac{P_{ns}^2}{(n^2/b^2+s^2/a^2)^2} - \frac{\varepsilon_i}{4\pi^4 a^2 D} \sum_{s=1}^{\infty} \left[ \sum_{n=1}^{\infty} \frac{s^2 P_{ns} \sin(n\pi x_i/b)}{(n^2/b^2+s^2/a^2)^2} \right]^2 \right. \\
\left. -\frac{\gamma_j}{2\pi^6 bD} \sum_{s,q=1}^{\infty} q \frac{s-q(-1)^{s+q}}{(s-q)(s+q)} \sum_{n=1}^{\infty} \frac{nP_{ns}\cos(n\pi x_i/b)}{(n^2/b^2+s^2/a^2)^2} \sum_{m=1}^{\infty} \frac{mP_{mq}\cos(m\pi x_i/b)}{(m^2/b^2+q^2/a^2)^2} \right\} \quad (5.4.5)$$

As before, if the expression in the curved brackets attains its global minimum for a certain $i$ at $x_i = x_0$, there is no reason why for different values of $i$, $x_i$ should not be equal $x_0$ if one wishes to find a global minimum point of $U$. This is true whenever,

$$U(\mathbf{x}) = \sum_{i=1}^{N} U_i(x_i) = \sum_{j=1}^{k}\sum_{i=1}^{N} U_i^j(x_i) = \sum_{i=1}^{N}\sum_{j=1}^{k} U_i^j(x_i) \quad (5.4.6)$$

and there is no additional equality constraint on the variables. Then the principle of separation of variables holds and the minimum of the sum is the sum of the minima (due to linearity, which always holds for a 1st order asymptotics). Then, all the minima are attained at the same point due to the fact that $U_i^j(x_i)$ is the same function of its argument for every $i$.

Hence, for flexible enough stiffeners – and small enough bending rigidity necessarily implies small enough torsional rigidity, since $E$ and $G$ are usually of the same order of magnitude and $I$ and $C$ for closed, thin-walled sections are both approximately proportional to the smaller dimension and to the third power of the larger dimension of the cross section – the optimal layout, in the since of minimal structural compliance, is a *single* stiffener. That is, there should be no splitting of stiffeners before a critical bending rigidity is supplied. This phenomenon was shown for bidirectional stiffening when neglecting torsional rigidity and for unidirectional stiffening when accounting for it. It is reasonable to believe that this should be at least approximately true for bidirectional stiffening with beams of non-negligible torsional rigidity (when the torsional rigidity is non-negligible with respect to the bending rigidity).

Moreover, in order to emphasize that the obtained result is not absolutely trivial, it should be noted that the 2$^{nd}$ order expansion in section 5.2 showed coupling between the variables implying that the objective function cannot be written as in (5.4.6). In the case examined there, it is not obvious that the stiffeners should not be split. However, subsection 5.2 proves that this is still true even if not by the argument shown in (4.6), but by a more delicate reasoning.

The combination of the results of sub-sections 5.1, 5.2, and 5.4 thus makes it not unreasonable to believe that for realistic bidirectional stiffening with closed cross-sections of non-negligible torsional rigidity, it may still be optimal to use only two orthogonal major stiffeners, for various loading distributions.

Beyond the abovementioned, numerical analysis of stiffened plates performed with the use of commercial software, showed that the effect of torsional rigidity, at least for the blade stiffeners examined in section 5.2, corresponds to no significant change in the total work done by the external loading. Numerical optimization also showed that when the optimal layout consists of one or two perpendicular stiffeners, it is true for various cross-sections, both when accounting for and neglecting the effect of torsional rigidity.

Last, in order to illustrate the result in (5.4.5), it should be pointed out that for a symmetric loading, $x_i = 1/2$ makes the 2$^{nd}$ term in (5.4.5) maximal in absolute value and the third term zero. This coincides with the understanding that for symmetric loading, the mid-span would not twist and therefore a stiffener placed there would apply no torque and torsional rigidity would not contribute to the strain energy of the structure.

# 6  CONCLUSIONS

This paper presented several results of an investigation in the subject of optimal stiffening of rectangular plates in elastostatic bending. A previously published method for the analysis of stiffened plate was obtained in more compact forms in sections 2, 3 and 4 herein and utilized to derive globally optimal stiffening layouts for limit-case parameters. It was shown that the optimal layout of each of two orthogonal sets of parallel, highly flexible stiffeners with negligible torsional rigidity consists of two orthogonal beams having increased rigidities positioned perpendicularly. In addition, the asymptotically globally optimal layouts for long stiffeners were obtained for both highly flexible and highly rigid stiffeners, to the 1$^{st}$ and 2$^{nd}$ orders of approximation for the former and the latter, respectively. The effect of the interaction of perpendicular stiffeners was shown to cancel out for stiffeners of negligible torsional rigidity. The effect of torsional rigidity was examined from the perspective of its ability to influence the optimal layout of asymptotically flexible stiffeners and was found to be insufficient, at least for a 1$^{st}$ order asymptotic expansion of the structural compliance with respect to the rigidities of the stiffeners, and for the special case of unidirectional stiffening. In addition, it should be mentioned that, although not shown here, the structural compliance minimization problem was solved for beams of different cross-sectional parameters, producing the result that the higher the bending rigidity of the stiffeners with respect to that of the plate, the more unidirectional the stiffening becomes in the optimum. Further noteworthy results correlating with the present work can be found in the works of Schade (1940), Clarckson (1965), Mróz and Rozvany (1975), Fuchs (1976), Samsonov (1978), Fuchs and Brull (1979), Cheng and Olhoff (1981), O'Leary and Harari (1985), Lam and Santhikumar (2003), Lagaros et al. (2004), Perchikov and Fuchs (2006) and Szczepanik (2006).

The essence of the contribution of this work to the existing body of knowledge is the utilization of a method of solution enforcing continuous interfaces between a plate and a set of stiffeners, including a derivation of a compact form of the strain energy of the structure, based on the method, for the purpose of the derivation of globally optimal positions of uniform stiffeners, at least for the asymptotic cases of highly flexible and highly rigid stiffeners and under the simplifying assumptions that the loading is harmonic and the stiffeners are at least relatively long. The obtained results seem to give insight into the structural behavior of such widely used structural elements as stiffened panels.

As globally optimal layouts are hard to obtain either in geometric or in topological optimization, and even more so when it comes to the determination of the symbolic dependence of the optimal layouts on the parameters, asymptotic analysis, together with a closed form of the solution of the elastostatic bending problem, are what it seems that might shed additional light on the area. Although asymptotic approach to the optimal stiffening problem has been employed before, the specific combination of assumptions and the results presented herein seem not too marginal to contribute to the understanding of the subject.

On another subject, the deflection field of a thin rectangular plate subjected to arbitrary boundary conditions and a general double-sine-series-expressible loading is given by Fletcher and Thorne (1955) and discussed by Goriupp (1947) and Timoshenko (1959). As the solution procedure given by Fletcher and Thorne produces the deflection field of a plate having arbitrary boundary conditions given the deflection field of a simply-supported plate subjected to a prescribed loading, clearly all one has to do is to supply the solution procedure given by these authors with the deflection field of a simply-supported *stiffened* plate instead of that of an *unstiffened* one.

In this paper, the effect of the torsional rigidity of the stiffeners was accounted for in a way similar to the way it is done by Mazurkiewicz (1962). More compact expressions for the entities representing the solution in the case where torsional rigidity is non-negligible can be obtained using analytic summation of the series in section 4 in the same manner in which it is done in sections 2 and 3 and in the work of Fletcher and Thorne (1955).

**Acknowledgments** The present work was partly supported by the Israeli Science Foundation under grant No. 13/05.